\documentclass[]{aa}  

\usepackage{graphicx}
\usepackage{txfonts}
\usepackage{natbib,twoopt}
\usepackage[breaklinks=true]{hyperref} 
\usepackage[switch]{lineno} 
\begin{document}

   \title{Tidal excitation of autoresonant oscillations  \\ 
   in stars with close-by planets}


   \author{A.~F.~Lanza
          }

   \institute{INAF-Osservatorio Astrofisico di Catania, Via S.~Sofia, 78 - 95123 Catania, Italy\\
              \email{antonino.lanza@inaf.it}
             }

   \date{Received XXX; accepted XXX}
\titlerunning{Autoresonance in star-planet systems}
\authorrunning{A.~F.~Lanza}
 
  \abstract
   {Close-by planets may excite various kinds of oscillations in their host stars through their time-varying tidal potential.}
   {Magnetostrophic oscillations with a frequency much smaller than the stellar rotation frequency have recently been proposed to account for the spin-orbit commensurability observed in several planet-hosting stars. In principle, they can be resonantly excited in  an isolated slender magnetic flux tube by a Fourier component of the time-varying tidal potential with a very low frequency in the reference frame rotating with the host. However, due to the weakness of such high-order tidal components, a mechanism is required to lock the oscillations in phase with the forcing for long time intervals ($10^{3}-10^{7}$~yr) in order to allow the oscillation amplitude to grow.}
   {We propose that the locking mechanism is an autoresonance produced by the non-linear dependence of the oscillation frequency on its amplitude. We suggest that the angular momentum loss rate is remarkably reduced in hosts entering autoresonance that contributes to maintain those systems in that  regime for a long time. }
   {We apply our model to a sample of ten systems showing spin-orbit commensurability and estimate the maximum drifts of the relevant tidal potential frequencies that allow them to enter the autoresonant regime. Such drifts are compared with the expected drifts owing to the tidal evolution of the planetary orbits and the stellar angular momentum loss in the magnetized winds finding that autoresonance is a viable mechanism in eight systems,  at least in our idealized model. }
   {The duration of the autoresonant regime and  the associated spin-orbit commensurability may be comparable with the main-sequence lifetimes of the host stars, indicating that gyrochronology may not be applicable to those hosts. }


   \maketitle
%

\section{Introduction}
Most of the extrasolar planets known to date  orbit close to their host stars because the most effective discovery techniques, the method of the transits and that of the radial velocity,  favor the detection of close-by planets. In particular, the probability of transiting a star of radius $R_{\rm s}$ for a planet on a circular orbit of radius $a$ is $R_{\rm s}/a$, while the amplitude of the radial-velocity wobble of the star is proportional to $a^{-1/2}$ \citep[e.g.,][]{CochranHatzes96}. As a consequence of the small star-planet separations, those systems are characterized by remarkable tidal interactions \citep[][]{Mathisetal13,Ogilvie14}. 

In a reference frame rotating with the star, the tidal potential due to a close-by planet is time independent only if the orbital plane coincides with the equatorial plane of the star, the orbit is circular, and the stellar rotation period is equal to the orbital period. In most of the known systems, at least one of such conditions is not verified and the time-dependent potential can excite waves in the fluid body of the star that are called {\em dynamical tide}. The Kepler space telescope has revealed several close stellar binary systems with eccentric orbits where dynamical tides produce multiperiodic oscillations of the optical flux. They have been called heartbeat stars \citep[e.g.,][]{Fuller17}. The phenomenology produced by close-by planets is much less extreme because their masses are on the order of $10^{-5}-10^{-3}$ of those of their host stars, nevertheless, the possibility of exciting inertial waves and/or gravity waves in the hosts has been recognized and analysed in several works \citep[e.g.][]{OgilvieLin07,Barker20,Ahuiretal21}. 

Recently, \citet{Lanza22} proposed a new component for the dynamical tide in late-type stars. Specifically, he suggested that resonant magnetostrophic oscillations with a frequency much smaller than the stellar spin frequency may be excited in slender magnetic flux tubes in the stellar interior by the tidal potential of a close-by planet the orbital period of which is in close commensurability  with the rotation period of the star. Owing to the very small amplitude of the  exciting tidal potential, such oscillations can grow only if they are kept in resonance with the exciting potential for very long time intervals on the order of $10^{3}-10^{7}$~years depending on the mass of the planet, the radius of the star, and the orbit semimajor axis. The required phase locking is produced by a non-linear dependence of the frequency of the oscillations on their own amplitude. The present work is dedicated to a detailed investigation of such a resonance-locking mechanism that represents a particular case of  {\em autoresonance}.

Autoresonance consists in the property of a wide class of non-linearly oscillating systems of self-adjusting the period and the phase of their oscillations in order to continuously extract energy from an external quasi-periodic forcing. Even in the presence of a slowly varying frequency of the external forcing, autoresonance keeps the oscillating system in phase with the forcing allowing a continuous growth of the oscillation amplitude. An introduction to autoresonance is provided by, e.g., \cite{Friedland09} and references therein. An application to celestial mechanics is presented by \cite{Friedland01}. 

In the present work, the autoresonant physical system is introduced in Sect.~\ref{model} together with a description of its relevant non-linearities. An application to the star-planet systems in  close spin-orbit commensurability considered by \citet{Lanza22} is presented in Sect.~\ref{applications}, while some consequences and the general relevance of the excited autoresonant oscillations in the host stars are discussed in Sect.~\ref{discussion}. 

\section{Model}
\label{model}

\subsection{Tidal potential in star-planet systems}
\label{tidal_potential_sect}
The gravitational tidal potential, $\Psi$, produced by a planet inside its host star is a solution of the Laplace equation $\nabla^{2} \Psi = 0$. Considering a reference frame with the origin in the barycenter of the star, the polar axis along its spin axis, and rotating with the angular velocity $\Omega_{\rm s}$ of the star, $\Psi$  can be expressed as \citep[e.g.,][Section~2.1]{Ogilvie14}
\begin{equation}
\Psi  =  
\Re \left\{ \sum_{l=2}^{\infty} \sum_{m=-l}^{l} \sum_{n=-\infty}^{\infty} \Psi_{lmn}  \right\}, 
\label{tidal_potential}
\end{equation}
where $\Re \{z \}$ is the real part of a complex quantity $z$ and the components of the Fourier series  of the potential are given by
\begin{equation}
\Psi_{lmn} = \frac{Gm_{\rm p}}{a} A_{lmn} (e,i) \left( \frac{r}{a} \right)^{l} Y_{l}^{m} (\theta, \phi)  \exp(j \omega_{mn} t), 
\label{tidal_potential_comp}
\end{equation}
where $j = \sqrt{-1}$, $G$ is the gravitation constant, $m_{\rm p}$ the mass of the planet, $a$ the semimajor axis of the relative orbit, $A_{lmn}(e,i)$ a coefficient depending on the eccentricity of the orbit $e$ and its obliquity $i$ \citep[that is, the angle between the stellar spin and the orbital angular momentum; see][]{Kaula61,MathisLePoncin09}, $r$ the distance from the center of the star of radius $R_{\rm s}$, $Y_{l}^{m}(\theta, \phi)$ the spherical harmonic of degree $l$ and azimuthal order $m$ (with $|m| \leq l$) that is a function of the colatitude $\theta$ measured from the polar (spin) axis and the azimuthal coordinate $\phi$, $\omega_{mn} \equiv (m\Omega_{\rm s}-n\Omega_{0})$, with $\Omega_{\rm s} \equiv 2\pi /P_{\rm rot}$ the rotation frequency of the star having a rotation period $P_{\rm rot}$ and $\Omega_{0} = 2\pi/P_{\rm orb}$ the orbital mean motion, where $P_{\rm orb}$ is the orbital period of the planet; $n$ is an integer specifying the time harmonic of the potential oscillation, and $t$ is the time. The series development given in Eq.~\eqref{tidal_potential_comp} is valid in the interior of the star, that is, for $r \leq R_{\rm s}$. 

The frequency of the magnetostrophic waves, $\omega$, that we shall consider in this work is much smaller than the stellar spin frequency $\Omega_{\rm s}$. They can resonate with a component of the tidal potential the frequency of which $|\omega_{mn}| = |m\Omega_{\rm s} - n\Omega_{0} | \ll \Omega_{\rm s}$, that implies $P_{\rm rot} / P_{\rm orb} \simeq n/m$. In other words, a close commensurability between the stellar rotation period and the orbital period of the planet is required for a tidal excitation of those waves. A sample of stars where such a commensurability between $P_{\rm rot}$ and $P_{\rm orb}$ has been observed is introduced and discussed by \cite{Lanza22}. 

Owing to the weakness of the resonant tidal potential component, $\Psi_{lmn}$, a long time interval is required to amplify the magnetostrophic oscillations up to a significant amplitude implying that the magnetic structure where they are excited must be very long lived (see~\ref{resonant_flux_tubes}).

\subsection{Resonant slender magnetic flux tubes}
\label{resonant_flux_tubes}
The interior structure of late-type stars consists of an inner radiative zone surrounded by a convective envelope. Magnetic fields are subject to a buoyancy force in the superadiabatically stratified layers of the convection zones, while they can reside for long times in the subadiabatic radiative zones because magnetic buoyancy becomes relevant only when the magnetized plasma comes in thermal equilibrium with the surrounding medium. The boundary layer between the radiative zone and the overlying convective envelope is characterized by some penetration of the convective motions into the stably stratified layers because the downwardly directed convective columns do not stop where the Schwarzschild criterion is no longer verified, but continue their motion below the convection zone due to their inertia. Such a boundary layer is characterized by a slightly  subadiabatic stratification and is called the overshoot layer by analogy with the case of the convective cores of upper main-sequence stars because of such undershooting convective motions. A description of the physical properties of the layer and of the convective downdrafts was originally given by \citet{Zahn91} and \cite{RieutordZahn95}. 

The overshoot layer allows the storage of strong azimuthal magnetic fields  thanks to its subadiabatic stratification parameterized by $\delta \equiv \nabla - \nabla_{\rm ad} < 0 $, where $\nabla$ is the gradient in the layer and $\nabla_{\rm ad}$ is the adiabatic gradient \citep[see][]{Kippenhahnetal12}. The  strong gradient of the rotation angular velocity observed in the Sun by means of helioseismic techniques \citep{Schouetal98}, makes the overshoot layer apt to  amplify toroidal fields so that it has been considered a possible seat for the solar and stellar hydromagnetic dynamos \citep[e.g.][]{Parker93}. Unfortunately, its physical properties are not precisely known because of the lack of a detailed description of the undershooting convection, the strong shear, and the possible molecular weight gradient there present \citep[e.g.,][]{SpiegelZahn92}. 

\begin{figure}
    \centering
    \includegraphics[width=9cm,height=6cm,angle=0,trim=1.9cm 2.0cm 2.0cm 2.0cm,clip]{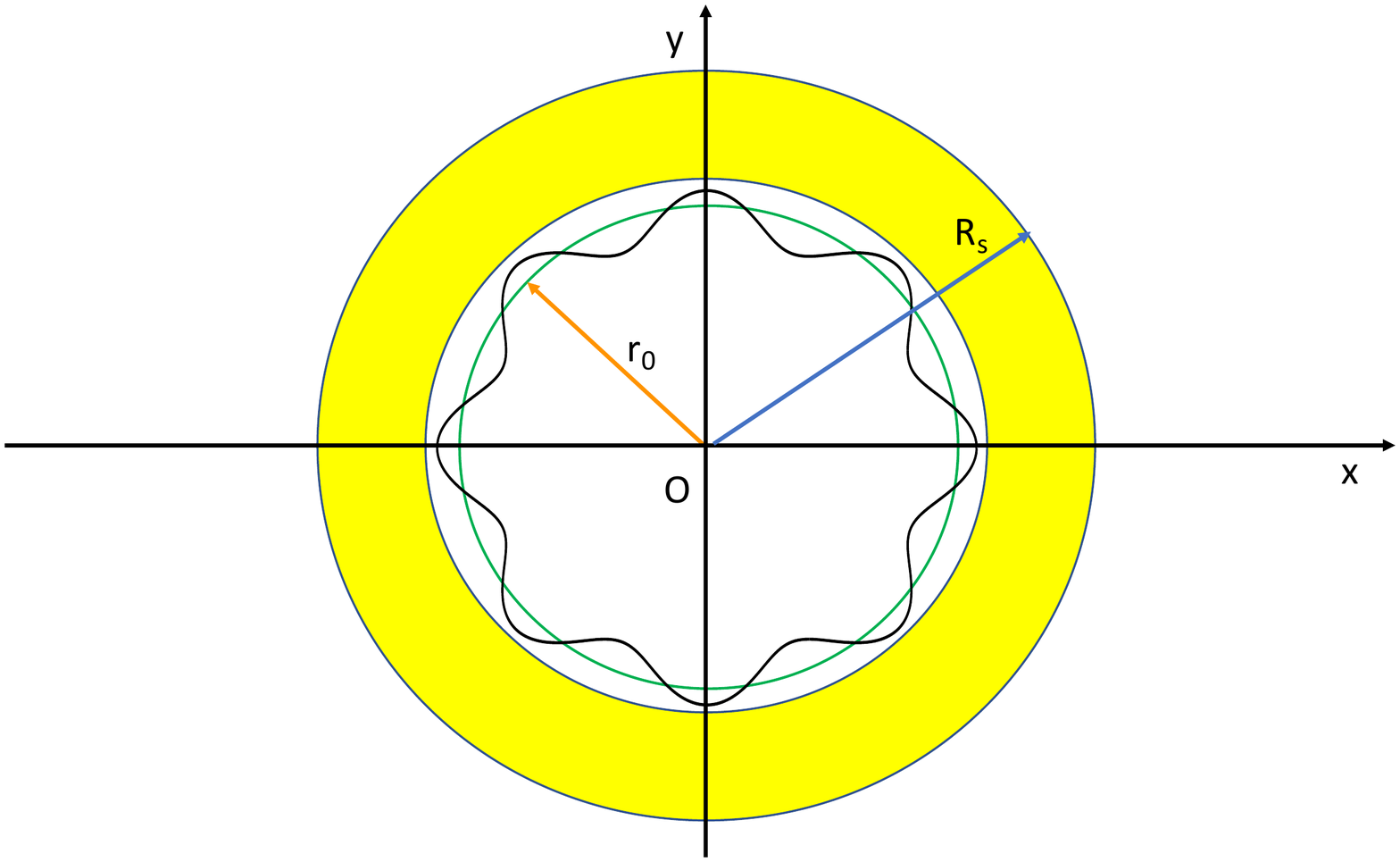}
    \caption{Sketch of a slender magnetic flux tube in equilibrium in the equatorial plane of a late-type main-sequence star (solid green line). The $xy$ plane is the equatorial plane of the star as seen from the North pole. The barycenter of the star is $O$, its radius is $R_{\rm s}$, while $r_{0}$ is the radius of the unperturbed flux tube. The convective zone of the star is rendered in yellow, while the radiative zone, including the overshoot layer where the flux tube is located, is rendered in white. The flux tube perturbed by an oscillation mode with azimuthal wavenumber $m=8$ is plotted as a black solid line. }
    \label{fig_flux_tube}
\end{figure}

A detailed treatment of the azimuthal magnetic fields  stored inside the overshoot layer is hampered by this lack of information, so that it is useful to model the azimuthal magnetic field as consisting of several slender azimuthal flux tubes separated by an unmagnetized plasma. For simplicity, we consider one of such axisymmetric flux tubes located in the equatorial plane of the star, so that its magnetic field is ${\vec B} = B {\vec e}_{\phi}$, where ${\vec e}_{\phi}$ is the unit vector in the azimuthal direction and $B$ is the intensity of the magnetic field (see Fig.~\ref{fig_flux_tube}).

We adopt the slender flux tube approximation that  assumes that the field is uniform across the section of the flux tube and that its section radius is much smaller than the curvature radius, assumed equal to the radius $r_{0}$ at the base of the convection zone, and the local pressure scale height, $H$. Such assumptions make the equations governing the motion of the slender flux tube particularly simple as shown by, for example, \citet{Ferriz-MasSchussler93} and \citet{Ferriz-MasSchussler94}. 

In our model, resonant magnetostrophic oscillations are excited in a slender toroidal flux tube and are slowly amplified along a timescale ranging from $\sim 10^{3}$ to $\sim 10^{7}$ years. This implies that the flux tube must remain stably stored inside the overshoot layer for those long time intervals, therefore, it is distinct from the flux tubes that emerge along the stellar activity cycle giving rise to active regions in the photosphere \citep{Caligarietal95}. The emergence of such slender flux tubes is produces by an undulatory instability that requires their magnetic field $B$ to exceed a threshold that depends mainly on the subadiabatic gradient $\delta$, the radius $r_{0}$, the acceleration of gravity $g$, and the stellar angular velocity $\Omega_{\rm s}$ \citep[e.g.,][]{SpruitvanBalle82,vanBalle83}. It is about $10^{5}$~G in the Sun and can reach several times $10^{6}$~G in rapidly rotating late-type stars with a deep convection zone \citep[cf.][]{Granzeretal00}. 

The magnetic field strength of the resonant flux tubes considered in our model is below the threshold for the onset of the undulatory instability being between $10$~G and a few times $10^{4}$~G in the systems investigated by \citet{Lanza22}. Therefore, such resonant flux tubes are stable to the undulatory instability and can be stored for long time intervals inside the subadiabatic overshoot layer. We assume that the field inside the flux tube is strong enough to overcome the effects of the convective downdrafts that tend to break it apart. If the velocity of the downdrafts, $\varv_{\rm d}$, is on the order of $10$ m~s$^{-1}$ in the layer where the tube is stored \citep[cf.][]{Zahn91}, this requires that the tube field $B \ga \varv_{\rm d} \sqrt{\mu \rho}$, where $\mu$ is the magnetic permeability of the plasma and $\rho$ the density in the layer. In the case of the Sun, this implies a field on the order of $100-500$~G that is much lower than the undulatory instability threshold. 

The overshoot layer can be the seat of a mild turbulence characterized by motions mainly in the horizontal direction because the stable stratification strongly hampers motions in the vertical (radial) direction. This can give rise to an anisotropic turbulent viscosity that can play a fundamental role in avoiding the spreading of the radial velocity gradient into the radiative zone \citep{SpiegelZahn92,BrunZahn06}. Nevertheless, by assuming that the magnetic field inside a resonant magnetic flux tube is larger than the equipartition field with the downdrafts and the associated turbulent motions, we can assume molecular values for the magnetic and thermal diffusivities across the flux tube and consider the molecular kinematic viscosity for the damping of the wave motions excited inside the flux tube itself \citep{Lanza22}. 

In view of such considerations, we assume the molecular values adopted by \citet{BrunZahn06} for the top of the solar radiative zone, that is,
a thermal diffusity $\kappa_{\rm T} = 10^{3}$~m$^{2}$~s$^{-1}$, a magnetic diffusivity $\eta_{\rm M} \sim 0.1$ m$^{2}$~s$^{-1}$, and a kinematic viscosity $\nu \sim 3\times 10^{-3}$ m$^{2}$~s$^{-1}$.  With a reference diameter of the resonant flux tube of $d_{\rm f} = 5 \times 10^{6}$~m, such values imply that the magnetic diffusion timescale of the flux tube is $d_{\rm f}^{2}/\eta_{\rm M} \sim 2.5 \times 10^{14}$~s, that is, on the order of $\sim 10^{7}$ years. In other words, the flux tube can have a very long lifetime, a necessary condition for the slow autoresonant wave excitation to operate. 
On the other hand, the thermal diffusion timescale is shorter by a factor of $10^{4}$, that implies that the resonant flux tube must be assumed in thermal equilibrium with its surroundings. The corresponding magnetic buoyancy can be counterbalanced by the aerodynamic drag force produced by the convective downdrafts, thus allowing the flux tube to be stably stored inside the lower part of the overshoot layer \citep[see Sect.~3.2 of][for more details]{Lanza22}. 

We stress that the considered resonant flux tubes do not take part in the stellar activity cycles having periods of a few decades. They should be regarded as the low-field tail of the magnetic flux tube distribution produced by the hydromagnetic stellar dynamo. They can be pumped into the lower part of the overshoot layer by the convective downdrafts and do not emerge to the surface, but are subject to a very slow decay under the action of their internal  molecular magnetic diffusivity. The processes producing slender flux tubes in the overshoot layers of the Sun and late-type stars are beyond the scope of this work and have been discussed, for example, by \cite{Schussler05} and \citet{Fan21}.  

\subsection{Tidally forced oscillations in stars}
\label{general_model}

{Our idealized model for the magnetic field configuration in the overshoot layer allows us to consider a  simple approach to study the tidal excitation of resonant oscillations in our model star. A more realistic model should consider the configuration of the internal magnetic field and the effects it has on the oscillation modes that propagate across the whole star adopting an approach similar to that of, e.g., \citet{Schenketal01}, \citet{LaiWu06},  \citet{Fuller17}, or  \citet{Lai21}. This approach shows that  the tidal excitation of a stellar oscillation mode requires not only a close match between the mode frequency and the frequency of a Fourier component of the tidal potential, but also a significant spatial overlap between the mode perturbation and the gradient of the exciting potential component as quantified by the so-called tidal overlap coefficient 
\begin{equation}
    {\cal Q}_{\alpha, lm} \equiv \int_{V} \rho({\bf r})\, \xi_{\alpha}^{*}\, ({\bf r}) \, \nabla\left( r^{l} Y_{l}^{m} \right) \, d^{3} {\bf r},
    \label{tidal_overlap_coeff}
\end{equation}
where $\rho ({\bf r})$ is the density of the plasma at position $\bf r$, $\xi_{\alpha}({\bf r})$ the mode spatial eigenfunction with $\alpha$ an index identifying the mode and the asterisk indicating complex conjugation, while  $r^{l} Y_{l}^{m}$ gives the spatial dependence of the potential of the exciting tidal component specified by the degree $l$ and the azimuthal number $m$; the integral is extended over the volume $V$ of the star \citep{LaiWu06}. 

In our simplified approach, the mode eigenfunctions coincide with the eigenfunctions of the slender flux tube oscillations. They are different from zero only inside the flux tube centered at $r=r_{0}$ in the equatorial plane of the star and have a spatial dependence proportional to $\exp (j m \phi)$ that overlaps  with the azimuthal dependence of the tidal potential component with the same azimuthal number $m$  (cf. sect.~\ref{forced_oscill_in_sft}, Eq.~\ref{xieq}). In other words, our highly idealized model allows an easy excitation of a slender flux tube mode provided that its frequency comes close to one of the frequencies of the tidal potential components and both the oscillation and the exciting tidal component have the same azimuthal wavenumber $m$ to warrant their spatial overlap. This is possible because we consider a single slender flux tube in isolation inside the star. Moreover, we assume the magnetic field and the density to be constant across the flux tube section to avoid any phase mixing that could damp the excited oscillations. Small gradients of the field or of the density can be tolerated if the flux tube magnetic field has a small twist as usually assumed in dynamical models to keep flux tubes coherent during their motion \citep[cf.][]{Fan21}. 

In a real star the situation can be rather different if several slender flux tubes can interact with each other. In this case, the tidal potential can excite oscillations only in the resonant flux tube, while no oscillations are excited in the neighbour flux tubes owing to the sharpness of the resonance (see Sect.~\ref{forced_oscill_in_sft}). Once the amplitude of the oscillations in the resonant tube becomes significant, its oscillations are strongly damped by magnetic reconnection with the unperturbed fields of the neighbour flux tubes whose oscillations are not excited. Therefore, our model can work only if the resonant flux tube is sufficiently isolated from neighbour flux tubes as to avoid any dynamical and magnetic interactions with them. Fortunately, such a condition is verified if the separation between neighbour flux tubes is at least of the order of $10^{6}$~m as we shall see in Sect.~\ref{auto_wind_tor} where we find that the strongest coupling is due to the turbulent viscosity in the overshoot layer, while magnetic diffusivity can be neglected in our approximation because the relatively strong fields of the flux tubes damp any turbulence inside them making the magnetic diffusivity equal to its molecular value, that is, much smaller than the turbulent viscosity.}

\subsection{Forced oscillations in a magnetic flux tube}
\label{forced_oscill_in_sft}

The equations describing the small free oscillations of a toroidal magnetic flux tube {around its equilibrium configuration} were derived by \citet{Ferriz-MasSchussler93} and \citet{Ferriz-MasSchussler94} who applied them to study the stability of the flux tube itself. A useful property consists in the decoupling of the oscillations in the equatorial plane from those in the latitudinal direction, that allows us to simplify our analysis {by considering only oscillation modes in the equatorial plane}. 
{We generalize Ferriz-Mas \& Sch{\"u}ssler's non-dimensional equations of motion  to the case of a flux tube subject to an external tidal force $-\rho_{0 i}\nabla \Psi$  per unit volume 
\begin{eqnarray}
\tau^{2} \left( \ddot{\xi}_{\phi} + 2\Omega_{\rm s} \dot{\xi}_{r} + \frac{1}{r_{0}}\frac{\partial \Psi}{\partial \phi_{0}}   \right)  & = & \nonumber \label{eq_motion1} \\ 
2 \frac{c_{\rm T}^{2}}{v_{\rm A}^{2}} f^{2} \frac{\partial^{2} \xi_{\phi}}{\partial \phi_{0}^{2}} + 4f \frac{c_{\rm T}^{2}}{v_{\rm A}^{2}} \left( f - \frac{1-x}{2\gamma} \right) \frac{\partial \xi_{r}}{\partial \phi_{0}},  \\ 
\tau^{2} \left( \ddot{\xi}_{r} - 2\Omega_{\rm s} \dot{\xi}_{\phi} + \frac{\partial \Psi}{\partial r}  \right)  & = & \nonumber \\ 
2 f^{2} \frac{\partial^{2} \xi_{r}}{\partial \phi_{0}^{2}} -4f \frac{c_{\rm T}^{2}}{v_{\rm A}^{2}} \left( f - \frac{1-x}{2\gamma} \right) \frac{\partial \xi_{\phi}}{\partial \phi_{0}} + T \xi_{r}, 
\label{eq_motion2}  
\end{eqnarray}
where  $\rho_{0i}$ is the internal density in the unperturbed flux tube,  a dot indicates the time derivative, $\vec \xi = (\xi_{\phi}, \xi_{r})$ is the small Lagrangian displacement in the equatorial plane with respect to the position of equilibrium; $\phi_{0}$ the azimuthal coordinate measured along the unperturbed flux tube;  $\tau$ a characteristic timescale defined as 
\begin{equation}
\tau \equiv \left( \frac{\beta H}{g_{0}} \right)^{1/2},
\label{tau_defin}
\end{equation}
with $\beta \equiv 2\mu p_{\rm i0}/B^{2}$, $p_{\rm i0}$ being the unperturbed internal pressure in the flux tube, $H$ the pressure scale height, $g_{0}$ the acceleration of gravity at the tube radius $r_{0}$; $c_{\rm T}$ is the tube wave speed
\begin{equation}
c_{\rm T} \equiv \frac{c_{\rm s} v_{\rm A}}{\sqrt{c_{\rm s}^{2} + v_{\rm A}^{2}}},
\end{equation}
with $c_{\rm s} = \sqrt{\gamma p_{\rm i0}/\rho_{\rm i0}}$ being the internal sound speed and $\gamma = 5/3$ the adiabatic exponent; 
 $f\equiv H/r_{0}$; $x \equiv  r_{0} \Omega_{\rm s}^{2}/g_{0}$; $x_{\rm e} \equiv r_{0} \Omega_{\rm e0}^{2}/g_{0}$; and $T$ is given in the limit $\beta \gg 1$ by
\begin{equation}
T \equiv 2 (\sigma -1) f^{2} + \frac{1}{\gamma} \left( 4f - \frac{2}{\gamma} +1 \right) + \beta \delta  - (1-\sigma) \tau^{2} \left( \Omega_{\rm e0}^{2} - \Omega_{\rm s}^{2} \right)  
- 4q \tau^{2} \Omega_{\rm e0}^{2},  
\label{T_eq}
\end{equation}
where $ \sigma \equiv \left[ d \log g(r) / d \log r \right]_{r_{0}} $ expresses the dependence of the acceleration of gravity on the radial distance and $q \equiv r_{0} \Omega^{\prime}_{\rm e}(r_{0})/[2 \Omega_{\rm e}(r_{0})]$ measures the shear due to the radial differential rotation at $r=r_{0}$ with the prime indicating the derivative with respect to the radial coordinate. 

The forcing produced by the mode of the tidal potential with frequency $\omega_{mn}$ can be written as: 
\begin{eqnarray}
\tau^{2} \frac{1}{r_{0}} \frac{\partial \Psi_{lmn}}{\partial \phi_{0}} = \hat{p}_{\phi} \exp (j\omega_{mn} t + jm \phi_{0}) \label{forcing1} \\
\tau^{2} \frac{\partial \Psi_{lmn}}{\partial r} = \hat{p}_{r} \exp (j\omega_{mn} t + jm \phi_{0}), 
\label{forcing}
\end{eqnarray}
where the complex quantities $(\hat{p}_{\phi}, \hat{p}_{r})$ can be deduced from the expression of the potential in Eq.~(\ref{tidal_potential_comp}), while the tidal frequency $\omega_{mn} = m \Omega_{\rm s} - n \Omega_{0}$ is real.  We shall consider only the stationary solution of the system of equations assuming that the effects of the initial conditions have had time to decay under the action of the viscous dissipation.  Alternatively,  the same solution is obtained if the oscillations are excited from an initial state of vanishing amplitude.

To solve Eqs.~(\ref{eq_motion1}) and~(\ref{eq_motion2}), we consider an individual mode of oscillation of the form
\begin{equation}
{\vec \xi} = \hat{\vec \xi} \exp (j\omega t + jm\phi_{0}), 
\label{xieq}
\end{equation}
where $\hat{\vec \xi}$ is a constant vector, $t$ the time, and the periodicity of the solution in the azimuthal direction imposes that $m$ be an integer (see Fig.~\ref{fig_flux_tube}).

Substituting Eqs.~(\ref{xieq}), (\ref{forcing1}), and (\ref{forcing}) into Eqs. (\ref{eq_motion1}) and (\ref{eq_motion2}), it is possible to reduce the latter to an algebraic system with $\omega = \omega_{mn}$ and the same $m$ of the component of the tidal potential \citep[cf.][ Eqs.~(7a) and (7b), for the corresponding freely oscillating system]{Ferriz-MasSchussler94}
\begin{equation}
\left( \begin{array}{cc}
A  &  -jD \\
jD & C  \\
\end{array} \right)
 \left( \begin{array}{c} \hat{\xi}_{\phi} \\ 
\hat{\xi}_{r} 
\end{array}
\right) = 
\left( \begin{array}{c} \hat{p}_{\phi} \\ 
\hat{p}_{r} 
\end{array}
\right),
\label{forced_osc_alg}
\end{equation}
where $\tilde{\omega}_{mn} = \tau\, \omega_{mn}$, $\tilde{\Omega}_{\rm s} = \tau \, \Omega_{\rm s}$,  
\begin{eqnarray}
A & \equiv & \tilde{\omega}_{mn}^{2} - 2 f^{2} m^{2}, \\
D & \equiv & 2 \left [\tilde{\Omega}_{\rm s} \tilde{\omega}_{mn} - 2 mf \left( f - \frac{1-x}{2 \gamma} \right) \right] ,  \mbox{and} \\
C & \equiv & A + T.   
\end{eqnarray}
The matrix of the algebraic system~\eqref{forced_osc_alg} is Hermitian. Therefore, we can diagonalize it and decouple the oscillations into two normal modes orthogonal to each other, each satisfying an equation
\begin{equation}
    \ddot{\Theta}_{i} + \omega_{mn}^{2} \Theta_{i} = P_{i}, 
\end{equation}
where $i = 1,2$ and $P_{i}$ are the components of the vector of the forcing terms~\eqref{forcing1} and~\eqref{forcing} after applying the coordinate transformation that leads to the normal modes (see below Eqs.~\ref{norm_mode1} and~\ref{norm_mode2}). 

In the present application, we focus on forced oscillations having a frequency much smaller than the stellar rotation frequency in the so-called magnetostrophic regime that is characterized by a flux tube magnetic field, $B$, such that the Alfven frequency is also much smaller than the rotation frequency. Following \citet{Ferriz-MasSchussler94}, such a condition can be written as $\varv_{\rm A} (m/r_{0}) \ll 2 \Omega_{\rm s}$, where $\varv_{\rm A} = B/\sqrt{\mu \rho}$ is the Alfven velocity along the magnetic flux tube, $m$ the azimuthal wavenumber of the oscillation mode and $r_{0}$ the radius of the flux tube. As shown in Sect.~4 of \citet{Lanza22}, such an approximation is well verified in the systems we shall investigate. 

For a magnetic field intensity on the order of a few hundreds Gauss, as considered in our model, $\beta \sim 10^{9}$ so that $|T| \sim \beta |\delta |\sim 10^{3}$ when we assume $|\delta | \sim 10^{-6}$, considered to be a typical value for the overshoot layer of the Sun or solar-like stars \citep[cf. Sect.~4 of][ or Table~\ref{table_auto_res_param}]{Lanza22}. This implies that $|A| \ll |T|$ in the limit $\omega_{mn} \ll \Omega_{\rm s}$, $f\sim 0.1$, and $ m\leq 10$. The eigenvalues $\Lambda_{i}$  of the matrix in Eq.~\eqref{forced_osc_alg}, each associated with the normal mode $\Theta_{i}$, verify $ |\Lambda_{1} | \ll |\Lambda_{2} |\sim |T|$, so that the amplitudes of the normal oscillation modes $\hat{\Theta}_{i}$ becomes 
\begin{eqnarray}
   \hat{\Theta}_{1} & = &  \left( \frac{\Lambda_{1}-C}{D} \right) \hat{\xi}_{\phi} + j \hat{\xi}_{r}   \simeq  j\hat{\xi}_{r}, \label{norm_mode1} \\
   \hat{\Theta}_{2} & = &  \left( \frac{\Lambda_{2}-C}{D} \right) \hat{\xi}_{\phi} + j \hat{\xi}_{r}   \simeq   \hat{\xi}_{\phi},  \label{norm_mode2}
\end{eqnarray}
where the final equalities imply a suitable normalization of the eigenvectors and we have made use of $A/D \sim \omega_{mn}/\Omega_{\rm s} \ll 1$ in the magnetostrophic regime. We shall explore the validity of these approximations from a quantitative viewpoint in Sect.~\ref{applications} for the eight systems investigated in the present study (cf. Table~\ref{normal_mode_table} for the values of the coefficients appearing in Eqs.~\ref{norm_mode1} and~\ref{norm_mode2}). 

The physical motivation for the almost orthogonal oscillations of $\xi_{\phi}$ and $\xi_{r}$ as  expressed by Eqs.~\eqref{norm_mode1} and~\eqref{norm_mode2} is the stable stratification of the overshoot layer that hampers any radial motion making the Coriolis force term $2\Omega_{\rm s} \dot{\xi}_{r}$ in Eq.~\eqref{eq_motion1} negligible. The stabilizing effect of such a  stratification is quantified by the large $|T|$. Therefore, the oscillations occur mainly in the azimuthal direction as already shown by the solutions of the equations of motion for the oscillations found by \citet{Lanza22} that imply $\hat{\xi}_{\phi} \sim |T|\, \hat{\xi}_{r}$. 
}

In view of the above considerations, we make the approximation $\Theta_{2} \sim \xi_{\phi}$, focussing on the oscillations of the toroidal flux tube in the azimuthal direction and writing its equation of motion in the form
\begin{equation}
    \ddot{\xi}_{\phi} + 2 \zeta \dot{\xi}_{\phi} + \omega^{2} \xi_{\phi} = \frac{1}{r_{0}} \frac{\partial \Psi_{lmn}}{\partial \phi_{0}},
\label{forced_oscill_eq}
\end{equation}
where $2 \zeta = \tau_{\rm d}^{-1}$ is the inverse of the viscous damping timescale given by $\tau_{\rm d} = d_{\rm f}^{2}/\nu$ with $d_{\rm f}$ the diameter of the flux tube and $\nu$ the kinematic viscosity of the plasma inside it as introduced in Sect.~\ref{resonant_flux_tubes}, while  $\omega$ is the  frequency of the magnetostrophic mode. {This frequency is given by the dispersion relation obtained by equating the determinant of the matrix in Eq.~\eqref{forced_osc_alg}  to zero because the damping term is so small that it has no practical effect on the eigenfrequencies of the free oscillations.  The presence of the small damping term avoids a singularity when the frequency of the forcing tidal potential becomes equal to the eigenfrequency of the magnetostrophic oscillations \citep[its expression is given in Eq.~29 of][and is not repeated here]{Lanza22}.

In principle, the thermal and Ohmic damping of the oscillations are also relevant because the molecular values of the thermal and magnetic diffusivities verify the inequality $\kappa_{\rm T} \gg \eta_{\rm M} \gg \nu$ in a stellar radiative zone. Nevertheless, the thermal and Ohmic diffusion timescales are on  the order of $\sim 10^{4}-10^{7}$ years (see Sect.~\ref{resonant_flux_tubes}), that is, they are comparable or longer than the typical growth timescales of the resonant oscillations  that range between $10^{3}$ and $10^{5}$ years (see Sect.~\ref{applications}). Therefore, we can neglect thermal and Ohmic damping of the oscillations in Eq.~\eqref{forced_oscill_eq} and consider only the smaller viscous damping because it gives the narrowest resonance width making their excitation the most difficult to achieve when the tidal frequency varies along with the evolution of the stellar spin and the orbital semimajor axis of the system.}

In the linear regime, we can neglect the dependence of the frequency of the magnetostrophic waves on their amplitude. However, such a dependence plays a fundamental role in the autoresonance process, therefore, we shall go beyond the linear regime  in the next subsection.

\subsection{Non-linear dependence of the oscillation frequency}
\label{oscill_freq_sect}
The frequency $\omega$ of the magnetostrophic oscillations in Eq.~(\ref{forced_oscill_eq}) depends on the magnetic field $B$ inside the oscillating flux tube. \citep{Lanza22} showed that it is a monotonously increasing function of $B$ and found that it depends on the amplitude of the oscillations through the perturbation of the mean magnetic field along the flux tube induced by the oscillations themselves. Specifically, he considered only the perturbation of the mean magnetic field induced by the radial displacement $\xi_{r}$ \citep[see Sect.~3.9 and, in particular, Eqs.~(45) and~(46) of][]{Lanza22} and found that it is capable of keeping the flux tube in resonance with the tidal potential in spite of small variations in its frequency $\omega_{mn}$ owing to tidal changes in the orbit of the planet and slow variations in the stellar rotation rate. 

In addition to the non-linear dependence of the oscillation frequency found by \citet{Lanza22}, two other dependences can be pointed out. The first depends both on the radial displacement $\xi_{r}$ and the  radial differential rotation and is introduced in Appendix~\ref{non-linear_dep_and_dr}. There we show that it is negligible in comparison to the second non-linearity, which depends on the azimuthal displacement $\xi_{\phi} \gg \xi_{r}$, and plays the most relevant role in the present model. Therefore, we introduce it in the following. 

The component of the momentum equation of the plasma along the instantaneous axis $\hat{\vec l}$ of the flux tube, that is, along its magnetic field lines, can be written as
\begin{equation}
 \rho \frac{\partial \varv_{l}}{\partial t} + \rho \, {\vec e}_{l}\cdot \left[ \left ({\vec \varv} \cdot \nabla \right) {\vec \varv} \right] = 
 -\frac{1}{r_{0}} \frac{\partial p}{\partial l} + \rho  \frac{\partial \Phi_{\rm G}}{\partial l}, 
 \label{azimuthal_mom_eq}
\end{equation}
where $\vec \varv $ is the velocity of the plasma inside the flux tube with $\varv_{l} = {\vec \varv} \cdot {\vec e}_{l}$ its component in the axial direction of unit vector ${\vec e}_{l}$, $t$ the time, $\rho$ the density, $p$ the pressure, and $\Phi_{\rm G}$ the gravitational potential. The component of the Lorentz force along the  flux tube axis is zero,  therefore, it does not appear in the right-hand side of Eq.~\eqref{azimuthal_mom_eq}. Given that the deformation of the flux tube is small ($\xi_{r} \ll \xi_{\phi}$), the deviation of the tube axis ${\vec e}_{l}$ from the unperturbed azimuthal direction ${\vec e}_{\phi}$ is a second-order effect when we regard the oscillation velocity field as a first-order quantity in the hypothesis of small oscillations. Therefore, we approximate $\varv_{l} \simeq \varv_{\phi}$, where $\varv_{\phi} = {\vec \varv} \cdot {\vec e}_{\phi}$ is the azimuthal velocity component, that is, the component along the axis of the unperturbed flux tube ${\vec e}_{\phi}$. In the same way, the derivative of any 
physical quantity along the direction ${\vec e}_{l}$ is approximated by its derivative in the azimuthal direction. In other words, we shall assume $\partial/\partial l \sim r_{0}^{-1} \, \partial / \partial \phi_{0}$, where $\phi_{0}$ is the azimuthal angle along the unperturbed flux tube. In particular, the 
stellar gravitational potential $\Phi_{\rm G}$ can be assumed to be axisymmetric,  thus allowing us to drop  the term  $\partial \Phi_{\rm G} / \partial l \sim r_{0}^{-1}\, \partial \Phi_{\rm G}/\partial \phi_{0}$ in Eq.~\eqref{azimuthal_mom_eq}. 

Equation~\eqref{azimuthal_mom_eq} can be put in a simpler form considering the  identity $\nabla (\varv^{2}/2) = \left ({\vec \varv} \cdot \nabla \right) {\vec \varv} + {\vec \varv} \times (\nabla \times {\vec \varv})$ and that the perturbations produced by the magnetostrophic oscillations are adiabatic because the thermal adjustment timescale of the flux tube is much longer than the oscillation period (cf. Sect.~\ref{resonant_flux_tubes}). This condition allows us to neglect the density perturbations $\delta \rho$ produced by the wave in comparison with the perturbation of the pressure $\delta p$ because they are smaller by a factor $\delta \rho/\delta p \sim \varv^{2}/c^{2}_{\rm s} \ll 1$, where $\varv$ is the wave velocity and $c_{\rm s}$ the sound speed in the layer.  

By applying the above results, we can recast Eq.~\eqref{azimuthal_mom_eq} in the form
\begin{equation}
    \rho \frac{\partial \varv_{\phi}}{\partial t} + \frac{1}{r_{0}} \frac{\partial}{\partial \phi_{0}} \left( \frac{1}{2} \rho \varv^{2} + p \right) = 0,   
    \label{bernoulli_eq}
\end{equation}
where the azimuthal component of the velocity is
\begin{equation}
     \varv_{\phi} = \varv_{0} + j\omega \, \hat{\xi}_{\phi} \exp (j\omega t + jm \phi), 
    \label{azimuthal_vel_eq}
\end{equation}
where $\varv_{0}$ is the uniform and constant azimuthal velocity that enters into the equation of the mechanical equilibrium of the unperturbed flux tube because the angular velocity of rotation of the plasma inside the flux tube is in general slightly different from that of the surrounding plasma \citep[see][Eq.~5]{MorenoInsertisetal92,Lanza22}. 

Substituing Eq.\eqref{azimuthal_vel_eq} into Eq.\eqref{bernoulli_eq} and taking into account that $\partial/\partial \phi_{0} \rightarrow j m$, we can integrate the resulting equation with respect to $\phi_{0}$ and obtain
\begin{equation}
 \frac{r_{0}\rho \omega}{m} \Re \left\{ \exp(j\omega t + jm\phi_{0}) \right\} + p + \frac{1}{2} \rho \varv_{\phi}^2 = f(t), 
 \label{eq_osc_vel}
\end{equation}
where $f(t)$ is a function of the time that plays the role of the integration constant that is independent of $\phi_{0}$ and we have approximated $\varv \simeq \varv_{\phi}$ because the radial component of the velocity is much smaller than the azimuthal component. Considering the nodes of the wave, that is, the points where its velocity is zero at any time $t$, we have
\begin{equation}
    f(t) = p_{0} + \frac{1}{2} \rho \varv_{0}^{2},
\end{equation}
where $p_{0}$ is the plasma pressure in the nodes that we assume equal to the uniform pressure inside the flux tube in equilibrium and in the absence of the wave. In this way, we have that $f(t) = F_{0}$ is a constant independent of the time $t$ and the azimuthal coordinate $\phi_{0}$. 

Making an average of Eq.~\eqref{eq_osc_vel} over  one oscillation period of the wave, the first order terms disappear and the average perturbation of the pressure inside the flux tube is found to be
\begin{equation}
    \langle \delta p \rangle + \frac{1}{4} \rho \omega^{2} \hat{\xi}_{\phi}^{2} = 0 , 
    \label{delta_p_eq}
\end{equation}
where $\delta p \equiv p -p_{0}$ is the pressure perturbation produced by the wave, while the angle brackets indicate the time average over one period of the sinusoidal oscillation. The lateral pressure balance across the flux tube is given by
\begin{equation}
    p + \frac{B^{2}}{2\mu} = p_{\rm e},
    \label{pressure_balance_eq}
\end{equation}
where $p_{\rm e}$ is the constant external pressure that is independent of $\phi_{0}$. Substituting Eq.\eqref{delta_p_eq} into Eq.\eqref{pressure_balance_eq}, we find the average variation of the magnetic field along the slender flux tube when a magnetostrophic wave of frequency $\omega$ is propagating along it
\begin{equation}
\Delta    \langle B^{2} \rangle = \frac{1}{2} \mu \rho \omega^{2} \hat{\xi}_{\phi}^{2}. 
    \label{mag_field_var}
\end{equation}
In the regime of magnetic field strength characteristic of our resonant flux tubes, the frequency of the magnetostrophic waves is linearly dependent on the average magnetic field, that is, $\Delta \omega/\omega = \langle \Delta B \rangle /B = (1/2) \Delta \langle B^{2} \rangle / \langle B^{2} \rangle $. Therefore, as a consequence of Eq.\eqref{mag_field_var}, the frequency of the wave varies non-linearly with its amplitude as
\begin{equation}
    \omega = \omega_{0} (1 + \gamma_{0} \hat{\xi}_{\phi}^{2}), 
    \label{wave_freq_eq}
\end{equation}
where 
\begin{equation}
    \gamma_{0} \equiv \frac{1}{4} \frac{\omega_{0}^{2}}{\varv_{\rm A}^{2}},  
\end{equation}
and $\omega_{0}$ is the frequency of the wave for a vanishing amplitude.  Neglecting the effect of the stratification on the frequency of the magnetostrophic mode, $\omega_{0} \approx \varv_{\rm A} m/r_{0}$, that gives the approximate relationship $\gamma_{0} \approx (1/4) (m/r_{0})^{2}$, independent of the magnetic field strength. The first-order perturbation analysis applied to derive Eq.\eqref{wave_freq_eq} is valid only when $\gamma_{0} \hat{\xi}_{\phi}^{2} \ll 1$, that is equivalent to $(\hat{\xi}_{\phi}/r_{0})^{2} \ll 1$ because of the above approximation. 

\subsection{Autoresonance}
\label{autoresonance_section}
Equation~\eqref{forced_oscill_eq} of the magnetostrophic oscillations, taking into account the non-linear dependence of their frequency on their amplitude found in Sect.~\ref{oscill_freq_sect} (cf. Eq.~\ref{wave_freq_eq}), can be written as
\begin{equation}
    \ddot{\xi}_{\phi} + 2 \zeta \dot{\xi}_{\phi} + \omega_{0}^{2} (1-\gamma_{\rm D} \xi^{2}_{\phi}) \xi_{\phi}= \frac{1}{r_{0}} \frac{\partial \Psi_{lmn}}{\partial \phi_{0}},
    \label{duffing_osc}
\end{equation}
where $\gamma_{\rm D} = -2\gamma_{0}$ and we have considered that $\gamma_{0} \hat{\xi}_{\phi}^{2} \ll 1$. Equation~\eqref{duffing_osc} is the equation of a forced Duffing oscillator with a small frictional damping \citep[e.g.][]{Wawrzynski21}, that is, one of the non-linear systems that shows  autoresonance \citep{Friedland09}. 

The dependence of the oscillation frequency on the amplitude shown by Eq.~\eqref{duffing_osc} allows it to adjust itself in order to remain in resonance with the external forcing, even when this is varying its frequency, provided that such a variation is sufficiently slow (see below). In such a way, the amplitude of the oscillations can continuously grow by extracting energy from the external forcing, even when its strength is very small because the resonance condition can be maintained for a time interval much longer than the oscillation period. 

A quantitative model of autoresonance can be formulated following different approaches developed to study non-linear oscillators \citep[e.g.,][]{Chirikov79,Friedland01}. We follow the multiple timescale method  as detailed in Appendix~\ref{autoresonance_eqs}. Defining a non-dimensional time $t^{\prime} \equiv \omega_{0} t$, the forced oscillator equation can be recast as
\begin{equation}
    \ddot{\xi}_{\phi} + 2 \lambda \dot{\xi}_{\phi} + \xi_{\phi} - \gamma_{\rm D} \xi_{\phi}^{3} = \eta \cos \varphi, 
\end{equation}
where $\lambda = \zeta \omega_{0}^{-1}$ and the forcing term in the right hand side of Eq.~\ref{duffing_osc}, depending on the tidal potential component $\Psi_{lmn}$, has been written as
\begin{equation}
    \frac{1}{\omega_{0}^{2} r_{0}} \frac{\partial \Psi_{lmn}}{\partial \phi_{0}} = \eta \cos \varphi,
\end{equation}
where $\eta > 0$ is the small amplitude of the oscillations of the potential the frequency of which is assumed to be slowly varying as
\begin{equation}
    \frac{\partial \varphi}{\partial t^{\prime}} = 1-\kappa \, t^{\prime}, 
    \label{kappa_eq}
\end{equation}
where $\kappa \, \omega_{0}$ is the inverse of the timescale of variation of $\omega_{mn} = m \Omega_{\rm s}- n \Omega_{0}$ due to the slow evolution of the orbit of the planet and the braking of the stellar rotation under the action of the stellar wind and the tides.  

The multiple timescale method seeks for a solution of the forced oscillator  equation of the form
\begin{equation}
    \xi_{\phi} = A \cos (t^{\prime} + \phi_{\rm A}),
\end{equation}
where $A$ is the slowly varying amplitude and $\phi_{\rm A}$ is the initial phase of the oscillation. The lag angle between the forcing and the oscillation can be defined as $\psi \equiv \theta - \varphi +\pi$, where $\theta \equiv t^{\prime} + \phi_{\rm A}$ is the phase of the oscillation. The oscillator is in resonance when $\psi = \pi/2$ implying that the forcing makes work to increase the amplitude of the oscillation during all the oscillation cycle. When the system is not too far away from resonance, $\psi$ is a slowly varying function of the time
such as the amplitude $A$. 

The multiple timescale method allows us to derive the equations describing the slow variations in $A$ and $\psi$ as (cf. Appendix~\ref{autoresonance_eqs})
\begin{eqnarray}
   \frac{dA}{d t^{\prime}} & = & -\lambda A + \frac{1}{2} \eta \sin \psi, 
   \label{aeq_evol}\\
   \frac{d\psi}{d t^{\prime}} & = & \kappa t^{\prime} - \frac{3\gamma_{\rm D}}{8} A^{2} + \frac{\eta}{2A} \cos \psi. \label{psieq_evol}
\end{eqnarray}
This system of ordinary differential equations can be numerically integrated to find the amplitude and the lag angle as functions of the time. Autoresonance corresponds to a regime where the angle $\psi$ is fixed or oscillating in an   interval centered on $\psi=\pi/2$ that corresponds to the resonance condition. In such a regime, the variations of $A$ and $\psi$ are characterized by oscillations that keep the system always in phase with the external forcing, thus allowing the mean amplitude of the oscillations to grow. For given values of the parameters $\gamma_{\rm D}$ and $\eta$, the autoresonant regime is possible only if $|\kappa|$ is smaller than a threshold value that depends on those parameters \citep{Friedland09}. 

To find the maximum value of $|\kappa| $ compatible with autoresonance, we follow the approach in Sect.~3 of \citet{Friedland01}. Let us assume that the system crosses the linear resonance keeping a continuous phase locking with an almost constant angle $\psi \ll 1$, so  that $\cos \psi \sim 1$ can be assumed to be  constant in Eq.~\eqref{psieq_evol}. We differentiate this equation with respect to $t^{\prime}$ and substitute Eq.~\eqref{aeq_evol} neglecting the very small dissipation term because $\lambda \ll 1$. In such a way, we obtain 
\begin{equation}
 \frac{d^{2} \psi}{dt^{\prime\, 2}} = \kappa - S \eta \sin \psi,
 \label{quasiparticle_eq}
\end{equation}
where 
\begin{equation}
    S \equiv \frac{3 \gamma_{\rm D}}{8} A + \frac{\eta}{4A^{2}}.
\end{equation}
Equation~\eqref{quasiparticle_eq}, describes the motion of a quasi-particle moving in a tilted cosine potential
\begin{equation}
    V_{\rm qp} (\psi) = -\left( \kappa \psi + \eta S \cos \psi \right). 
\end{equation}
In other words, the time variation of the angle $\psi$ is described by the unidimensional motion of a point mass ruled by the Lagrangian ${\cal L} = (1/2) \dot{\psi}^{2} - V_{\rm qp} (\psi)$. The system is in autoresonance if and only if the angle $\psi$ can oscillate around a fixed point $\psi_{0}$ that corresponds to a minimum of the potential $V_{\rm qp}$. This implies that autoresonance is possible only if the potential has a relative minimum where $d V_{\rm qp}/ d\psi = 0$, in other words 
\begin{equation}
    \kappa - \eta S \sin \psi = 0 
\end{equation}
or 
\begin{equation}
   | \kappa | = \eta S  |\sin \psi|  \leq \eta S, 
    \label{alpha_autores}
\end{equation}
otherwise the potential $V_{\rm qp}$ has no relative minima and increases monotonically leading to a monotone variation of $\psi$ in time. The  function $S(A)$ has a minimum at the point where $dS/dA = 0 $, that is, at $A = A_{\rm min}= [(4/3)(\eta/\gamma_{\rm D})]^{1/3}$. The corresponding minimum value of $S$ is $S_{\rm min} \equiv S(A_{\rm min})$ and it can be substituted into Eq.\eqref{alpha_autores} to obtain the condition to be verified by $\kappa$ for the system to enter into an autoresonant regime, that is, 
\begin{equation}
   | \kappa | \leq \kappa_{\rm th}
    \label{kappa_cond}
\end{equation}
where the threshold value $\kappa_{\rm th}$ is given by  
\begin{equation}
    \kappa_{\rm th} \equiv \eta S_{\rm min} = \left( \frac{3}{4}\right)^{5/3} \eta^{4/3} |\gamma_{\rm D}|^{2/3}.
   \label{kappa_thres}
\end{equation}
The value of $\kappa_{\rm th}$ given by Eq.~\eqref{kappa_thres} is an order-of-magnitude estimate because of the simplifying assumption adopted to derive Eq.~\ref{quasiparticle_eq}, but it is adequate for our purposes in view of the comparable or larger uncertainties in the current tidal and magnetic braking effects in planet-hosting stars (cf. Sect.~\ref{frequency_var_general}).

{External perturbations acting on the flux tube may produce a phase shift of its oscillations and destroy the resonance locking. An order of magnitude estimate based on the energy integral of the equation of the motion \eqref{quasiparticle_eq} requires that the maximum value of the non-dimensional time derivative of the phase lag verifies $\dot{\psi}_{\max} \la (2 k_{\rm th})^{1/2}$ to keep the oscillations of $\psi$ confined around a minimum of $V_{\rm qp}$. This translates into a condition on the characteristic timescale of any external perturbation $t_{\rm pert}$ that must be $t_{\rm pert} \ga (2 k_{\rm th})^{-1/2} \omega_{0}^{-1}$, that is, the perturbation must be sufficiently slow to allow the system to maintain the phase locking. Considering as typical values $k_{\rm th} \sim 10^{-7}$ and $\omega_{0} \sim 5 \times 10^{-8}$~s$^{-1}$ (see Sect.~\ref{applications}), we find $t_{\rm pert} \ga 1400$ years, that is a very long timescale in comparison to those of the known perturbations that can act on the flux tube. Therefore, the effects of such perturbations cannot be neglected and should be included as an additional term on the right-hand side of Eq.~\eqref{psieq_evol}. Nevertheless, the oscillation period of the lag angle $\psi$ in the absence of external perturbations ranges between $10^{2}$ and $10^{5}$~years as we shall see in Sect.~\ref{applications}. The known perturbations are periodic with a much shorter period, for example, the deformation due to the $l=2$ component of the tidal potential has a frequency $2(\Omega_{0} - \Omega_{\rm s})$, or are random with typical timescales of $\sim 10^{6}-10^{7}$~s as in the case of the convective downdrafts. Therefore, we can average the additional term  accounting for those external perturbations in Eq.~\eqref{psieq_evol} over a timescale remarkably longer than the perturbation  timescales, yet much shorter than the period of the unperturbed oscillations of $\psi$, making its contribution vanish. In other words, the large difference between the timescale of the $\psi$ oscillations and those of the known perturbations  acting on the flux tube makes it possible to neglect the effects of the latter on the autoresonant excitation of the former. }

The asymptotic increase of the mean amplitude in autoresonance can be derived by considering that the term proportional to $A^{-1}$ in the right-hand side of Eq.~\eqref{psieq_evol} becomes negligible in the asymptotic regime, while $d\psi/dt^{\prime} \sim 0$ on the left-hand side because the angle $\psi$ makes small oscillations around a minimum of the pseudopotential. Therefore,  we obtain the asymptotic scaling $A^{2} (t^{\prime}) \sim (8/3) \kappa t^{\prime}/\gamma_{\rm D}$. Considering two time instants $t^{\prime}_{0}$ and $t^{\prime}$, this gives
\begin{equation}
    A(t^{\prime}) \simeq \left[ A_{0}^{2} + \frac{8\kappa}{3 \gamma_{\rm D}} \left( t^{\prime} - t^{\prime}_{0} \right) \right]^{1/2},
    \label{asymptotic_sol}
\end{equation}
where $A_{0}\equiv A(t^{\prime}_{0})$ is the value of $A$ at the initial time $t^{\prime}_{0}$ when the autoresonant asymptotic regime can be assumed to begin. 

Formally, the asymptotic solution given by Eq.~\eqref{asymptotic_sol} is valid until the system reaches its  maximum amplitude $\hat{\xi}_{\phi \max}$ when the external forcing is balanced by dissipation, that is, $\hat{\xi}_{\phi \max} = A_{\rm max} \sim \eta \sin \psi /( 2\lambda)  \leq \eta /(2\lambda) $ (cf. Eq.~\ref{aeq_evol}). It corresponds to a great amplification of the forcing amplitude because  $\lambda^{-1} = \zeta^{-1} \omega_{0}$ is large, thus allowing  the small tidal potential oscillations to produce a sizeable effect as already discussed by \citet{Lanza22}. 

In the high-Reynolds number environment of the overshoot layer, the oscillations are likely to produce turbulence that limits their amplitude in the asymptotic regime. Inside the flux tube, the magnetic field can effectively dump any turbulent motion, therefore, turbulence will not directly limit the $\xi_{\phi}$ displacement, but the radial displacement $\xi_{r}$ is affected by the turbulence in the surrounding medium produced by the oscillations themselves. Such a turbulence  can effectively damp $\xi_{r}$, if its amplitude grows too large. The maximum amplitude achievable in that case has been estimated in Sect.~3.7, Eq.~(39), of \citet{Lanza22} and can be expressed in our formulation as $\hat{\xi}_{r \max} \sim \omega_{mn}^{-2} (\partial \Psi_{lmn}/\partial r)$. The maximum azimuthal amplitude $\hat{\xi}_{\phi \max}$ is linked to the maximum radial amplitude by $\hat{\xi}_{\phi \max} \sim |T|\, \hat{\xi}_{r  \max}$ because of the equations of motion governing the flux tube oscillations (cf.~Eqs.~\ref{eq_motion1} and~\ref{eq_motion2} and Sect.~\ref{forced_oscill_in_sft}), so that turbulence in the surrounding medium will limit also the azimuthal displacement. 

In some cases, the value of $\hat{\xi}_{\phi \max}$ obtained from the above conditions is so large that the small-displacement approximation $|\gamma_{\rm D}|\, \xi_{\phi}^{2} \ll 1$ of our first-order theory is no longer valid. When this happens, we shall conservatively assume a maximum $\hat{\xi}_{\phi\, \rm max} = 0.1 | \gamma_{\rm D}|^{-1/2}$ in such a way that the small-displacement approximation remains valid with $|\gamma_{\rm D} | \, \xi_{\phi}^{2} \leq 0.01$. In summary, the value of $\hat{\xi}_{\phi \max}$ is computed as the minimum between {\it i) }the equilibrium amplitude as given by Eq.~\eqref{aeq_evol}, that is, $\eta/(2\lambda)$; {\it ii)} the asymptotic amplitude as given by Eq.~\eqref{asymptotic_sol} with $\kappa = \kappa_{\rm th}$ and $t=10^{7}$~yr corresponding to the diffusion timescale of the resonant flux tube; {\it iii)} the maximum amplitude in the presence of self-generated turbulence; and {\it iv)} the requirement of small oscillations, that is, $\hat{\xi}_{\phi \max} \leq 0.1 |\gamma_{\rm D}|^{-1/2}$. The value of $\hat{\xi}_{\phi \max}$ determined in this way has to be regarded as an order-of-magnitude estimate in view of the uncertainties on $\kappa_{\rm th}$, $\lambda$, and the diffusion timescale of the resonant flux tubes, or the possibility of adopting a less strict smallness condition on $|\gamma_{\rm D}| \xi_{\phi}^{2}$.

\subsection{Frequency variation in the exciting tidal potential $\omega_{mn}$}
\label{frequency_var_general}
The autoresonance regime can be accessed only if the drift of the excitation frequency $|\kappa|$ is sufficiently small as shown by Eq.~\eqref{kappa_cond}. The sharpness of the resonance peak due to the smallness of the dissipation parameter $\lambda$ \citep[see][]{Lanza22}, implies that $\omega_{0}$ is virtually equal to  $\omega_{mn}$ at resonance. In other words,  $\kappa$, as defined in Eq.~\eqref{kappa_eq}, is given by 
\begin{equation}
    \kappa = -\frac{1}{\omega_{mn}} \frac{d\omega_{mn}}{dt^{\prime}} = -\frac{1}{\omega_{mn}^{2}} \frac{d\omega_{mn}}{dt}.  
    \label{kappa_vs_omega_mn}
\end{equation}
The variation in the frequency of the forcing tidal potential $\omega_{mn}$, responsible for the excitation of the oscillations, can be written as
\begin{equation}
    \frac{d \omega_{mn}}{dt} = m \frac{d \Omega_{\rm s}}{dt} - n \frac{d\Omega_{0}}{dt}, 
    \label{omega_mn_dot}
\end{equation}
where the variation in the angular velocity of the stellar rotation is produced by the action of the stellar magnetized wind and the tides raised by the planet, while the variation in the orbital frequency (mean motion) is due to the tides. Specifically, adopting the tidal model of \citet{MardlingLin02}, we can write
\begin{equation}
  I_{\rm s}  \frac{d\Omega_{\rm s}}{dt} = \frac{dL_{\rm W}}{dt} + \frac{9}{2Q^{\prime}_{\rm s}} \left(\frac{Gm_{\rm p}^{2}}{a} \right) \left( \frac{R_{\rm s}}{a}\right)^{5} \left( \frac{\Omega_{0}- \Omega_{\rm s}}{\Omega_{0}}\right), 
  \label{star_rot_evol_eq}
\end{equation}
where $I_{\rm s}$ is the moment of inertia of the star, $dL_{\rm W}/dt$ the angular momentum loss rate due to the stellar wind, $Q^{\prime}_{\rm s}$ the stellar modified tidal quality factor, $G$ the gravitation constant, $m_{\rm p}$ the mass of the planet, $a$ the orbit semimajor axis, and $R_{\rm s}$ the star radius. The moment of inertia is scaled from the solar value because the planet hosts we consider have an internal stratification similar to that of the Sun, that is, $I_{\rm s}= 0.07 M_{\rm s}\,  R_{\rm s}^{2}$, where $M_{\rm s}$ is the mass of the star. When $\Omega_{\rm s}$ is a function of the position inside the star, the stellar angular velocity entering into the specification of the tidal torque is the angular velocity of the convection zone because the tidal dissipation in the radiative interior is remarkably smaller due to the smaller radius of the radiative zone \citep[cf.][]{OgilvieLin07,Barker20}. 

The angular momentum loss rate due to the stellar wind that appears in Eq.~\eqref{star_rot_evol_eq} can be parameterized as
\begin{equation}
    \frac{dL_{\rm W}}{dt} = \left\{ \begin{array}{cc}
       -K_{\rm W} \Omega_{\rm s}^{3} (R_{\rm s}/R_{\odot})^{1/2} (M_{\rm s}/M_{\odot})^{-1/2} \mbox { for $\Omega_{\rm s} < \Omega_{\rm sat}$,}   \\
       -K_{\rm W} \Omega_{\rm s} \Omega_{\rm sat}^{2} (R_{\rm s}/R_{\odot})^{1/2} (M_{\rm s}/M_{\odot})^{-1/2}   \mbox{ for $\Omega_{\rm s} \geq \Omega_{\rm sat}$,}\\
    \end{array} \right. 
    \label{angular_mom_loss_rate}
\end{equation}
where $K_{\rm W}=2.7 \times 10^{40}$~kg~m$^{2}$~s$^{-1}$ is a constant calibrated with the present Sun, $\Omega_{\rm sat}$ is the saturation angular velocity that is a function of the mass of the star \citep{Bouvieretal97}. We shall assume $\Omega_{\rm sat} = 5, 9, 14, 16$~$\Omega_{\odot}$ for a star mass $M_{\rm s} = 0.5, 0.8, 1.0, 1.2$~M$_{\odot}$, respectively, and linearly interpolate for masses in between.

Stars with close-by giant planets may experience a reduced angular momentum loss rate because a hot Jupiter could reduce the Alfven radius of the stellar wind and the mass loss rate by modifying  the large-scale configuration of the coronal field. The decrease in the angular momentum loss rate may reach two orders of magnitude with respect to  Eq.~\eqref{angular_mom_loss_rate}, depending on the parameters of the system \citep{Cohenetal10,Lanza10,Lanza14}. 

The modified tidal quality factor of the star, $Q^{\prime}_{\rm s}$, depends on the processes responsible for the dissipation of the tidal kinetic energy inside the star. When the rotation period is shorter than twice the orbital period, tidal inertial waves can be excited in a main-sequence star leading to a remarkable increase of the tidal dissipation that corresponds to a decrease of $Q^{\prime}_{\rm s}$ with respect to the regime where $P_{\rm rot} > 2 P_{\rm orb}$ in which tidal dissipation is produced only by the equilibrium tide \citep{OgilvieLin07}. Following  \citet{CollierCameronJardine18}, we assume $\log_{10} Q^{\prime}_{\rm s} = 7.3$ when inertial waves can be excited and $\log_{10} Q^{\prime}_{\rm s} = 8.26$ when only the equilibrium tide is present. 

The variation in the orbital mean motion, $\Omega_{0}$, due to the tides, is given by
\begin{equation}
    \frac{1}{\Omega_{0}} \frac{d\Omega_{0}}{dt} = \frac{27}{2Q_{\rm s}^{\prime}} \left( \frac{m_{\rm p}}{M_{\rm s}} \right) \left( \frac{R_{\rm s}}{a}\right)^{5} (\Omega_{0} - \Omega_{\rm s}), 
    \label{mean_motion_var}
\end{equation}
where the symbols have already been introduced above.

Neglecting the tides and the possible reduction in the Alfven radius due to a close-by massive planet, the rate of angular momentum loss in the stellar wind would in several cases produce a value of $|\kappa| $ larger than the maximum allowed by Eq.~\eqref{kappa_cond} when the frequency $\omega_{mn}$ crosses the resonance. The discrepancy is by at least one or two orders of magnitude in several systems (cf. Sect.~\ref{applications}). Therefore, we must invoke the effects of tides raised by the planet or effects not included in the above simplified model to account for the initial locking of those star-planet systems into autoresonance. 

A necessary condition for tides to counteract the magnetic braking is that $\Omega_{0} > \Omega_{\rm s}$ because only in such a case they transfer angular momentum from the orbit to the stellar spin counteracting the angular momentum loss produced by the stellar wind. Conversely, when $\Omega_{\rm s} > \Omega_{0}$  tides contribute to brake the star by transferring angular momentum from the stellar spin to the orbit (cf. Eq.~\ref{star_rot_evol_eq}). Moreover,  tidal effects can be significant only when the planet has a mass at least comparable with that of  Jupiter  and a relative orbital separation $a/R_{\rm s} \la 10$. When $a/R_{\rm s} \ga 10$, the tidal torque on the stellar spin is too small, even for a very massive hot Jupiter ($ m_{\rm p} \sim 10$~M$_{\rm J}$) and assuming a reduced Alfven radius, because the torque  scales as $(a/R_{\rm s})^{-5}$ (cf. Eq.~\ref{mean_motion_var}) and the minimum value we can adopt for $Q_{\rm s}^{\prime} \sim 10^{6}$ for the late-type main-sequence hosts under consideration  \citep[e.g.,][]{OgilvieLin07,Barker20}. 

In the regime where tides due to the planet are not efficient, the variation in $\omega_{mn}$ is dominated by the evolution of the stellar spin. Therefore, the possibility of making $|\kappa|$ smaller than the threshold value for triggering autoresonance could be related to a temporary stalling of the evolution of the stellar spin. Since the flux tube oscillations predicted by our model require a timescale shorter than $10^{6}-10^{7}$~yr for their amplification, a stalled rotation phase with a duration much shorter than the typical timescale of stellar spin evolution ($10^{8}-10^{9}$~yr) is sufficient to allow the system to enter into autoresonance. Once established, the autoresonant regime can maintain itself for long time intervals, in some cases comparable with the main-sequence lifetime of the system itself, as discussed by \citet{Lanza22} and as we shall show in Sect.~\ref{evolution_of_omega_mn}. 

Observations of stellar rotation in open clusters in the age range $\approx 0.7-2.7$~Gyr suggest that a stalling of the rotational evolution can indeed occur for main-sequence stars with $M_{\rm s} \la 0.8$~M$_{\odot}$ with its beginning and duration principally depending on the stellar mass \citep[e.g.,][]{Aguerosetal18,Curtisetal19,Douglasetal19}. Specifically, while the rotational evolution of stars with a mass of $\sim 0.8$~M$_{\odot}$ can be stalled for $\approx 0.3$~Gyr, their lower mass syblings of $\approx 0.55$~M$_{\odot}$ appear to be stalled for $\sim 1.3$~Gyr \citep{Curtisetal20}. Stars with a mass higher than 0.8~M$_{\odot}$ could also experience a phase of stalled rotation, but it may have gone unnoticed because of a duration on the order of $30-100$~Myr, that is, shorter than the shortest separation in age between the clusters analyzed to detect the phenomenon.  

\citet{SpadaLanzafame20} interpreted the stalled rotation phase by invoking an exchange of angular momentum between the radiative interior and the convective envelope in late-type main-sequence stars. Their two-zone model predicts a phase when the resurfacing of angular momentum stored into the radiative interior can counteract the braking of the convection zone by the stellar wind, temporarily stalling the rotational evolution of  a star. During such a phase, the value of $|\kappa|$ may become small enough to lock the system into autoresonance. This will lead to a significant amplification of the flux tube oscillations in the systems for which one of the tidal potential frequencies crosses the oscillation frequency of one of the slender flux tubes in the overshoot layer. Given that the magnetic field intensity of the potentially resonant flux tubes has a rather wide distribution, the possibility for one of the frequencies of the tidal potential components to come into resonance with one of the flux tubes is not a remote one. The observed duration of the phase of stalled rotation, ranging between $\sim 0.3$ and $1.3$~Gyr depending on the star mass, is certainly much longer than the amplification timescale of the autoresonance process, which makes it even more likely to be triggered (cf. Sect.~\ref{applications}).

\subsection{Wind torque in autoresonant systems}
\label{auto_wind_tor}
Once the autoresonant oscillations are excited and grow to a significant amplitude, the azimuthal velocity field of the wave inside the flux tube propagates in its immediate neighbourhood. The propagation of the oscillating azimuthal flow is ruled by the Navier-Stokes equation because the medium around the flux tube is unmagnetized. The radial component of the oscillation velocity field is on the order of $|T|^{-1} \sim 10^{-3}$ of the azimuthal component and is in general smaller than the velocity field of the convective downdrafts that is on the order of $10-100$~m~s$^{-1}$ in the mid of the overshoot layer \citep{Zahn91,RieutordZahn95}. The downdrafts are braked and their flow is turned horizontal close to the base of the overshoot layer. Their velocity field is a random field with a coherence timescale on the order of $10^{6}-10^{7}$~s, corresponding to the convective turnover time close to the base of the overlying convection zone, and a spatial coherence scale on the order of $10^{7}$~m or smaller \citep[cf. Eq.~13 and Sect.~3.2 in][]{Pinconetal21}. On the other hand, the azimuthal component of the oscillation velocity field has typical coherence timescales and lengthscales  one order of magnitude larger than those of the downdraft field and is periodic (cf. Tables~\ref{table_auto_res_param} and~\ref{table_tidal_freq_param} in Sect.~\ref{applications}). Therefore, it  will dominate the development of the undulatory instability in the neighbour flux tubes. 

Specifically, the coherently oscillating velocity field of an autoresonant flux tube may trigger the development of the undulatory instability in neighbour flux tubes with a field strength $\geq 10^{5}$~G that are regarded to be responsible for the formation of active regions at the surface of the star \citep{Caligarietal95}. The azimuthal wavenumber of the most unstable mode is $m=1$ or $m=2$ \citep{Ferriz-MasSchussler94,Granzeretal00}, therefore, the growth of the instability can be initiated close to one of the nodes of the azimuthal velocity field of a neighbour resonant flux tube because its azimuthal velocity field propagates  to the flux tube whose magnetic field is close to the threshold for the development of the undulatory instability. {The amplitude of the perturbation does not need to be large because the unstable state of the strong-field flux tube will exponentially amplify it with a characteristic growth timescale on the order of $10^{7}$~s \citep{Caligarietal95,Schussler05,Fan21}. This allows the resonant flux tube to be still considered as virtually isolated from the strong-field flux tube and its oscillations are not damped by magnetic reconnection as discussed in Sect.~\ref{forced_oscill_in_sft} also thanks to their very small radial displacement  $\xi_{r} \sim |T|^{-1} \xi_{\phi} \sim 10^{-3} \xi_{\phi}$. }

The propagation of the azimuthal velocity field into the fluid around an oscillating flux tube is due to the small viscosity present in the overshoot layer. Assuming that the propagation lengthscale is small in comparison with the radius $r_{0}$  and the diameter of the cross section of the oscillating flux tube, $d_{\rm ft}$, the Navier-Stokes equation can be approximated as 
\begin{equation}
    \frac{\partial \varv_{\phi}}{\partial t} \simeq \nu \frac{\partial^{2} \varv_{\phi}}{\partial r^{2}},
    \label{ns_eq}
\end{equation}
where $\nu$ is the kinematical viscosity in the overshoot layer. We seek for a solution of the kind $\varv_{\phi} (r, \phi, t) \propto \xi_{\phi} (\phi, t) F(r)$, assuming a radial dependence $F(r) \propto  \exp[-(1+j) \lambda_{\rm F} r]$, where $\lambda_{\rm F}$ is the inverse lengthscale ruling the radial propagation of the flow in the neighbourhood of the oscillating flux tube \citep[cf.][ Sect.~4.3]{Batchelor68}. By substituting such an anzats into Eq.~\eqref{ns_eq}, we find the inverse propagation lengthscale 
\begin{equation}
    \lambda_{\rm F} = \left( \frac{\omega_{mn}}{2\nu} \right)^{1/2}. 
\end{equation}
Considering a frequency of the tidal potential in the rotating reference frame of the star $\omega_{mn} \sim 10^{-4} \Omega_{\rm s} \sim 10^{-10}$~s$^{-1}$ and a molecular viscosity $\nu \sim 3 \times 10^{-3}$~m$^{2}$~s$^{-1}$, the propagation lengthscale $\lambda_{\rm F}^{-1}$ is smaller than $10^{4}$~m. Nevertheless, considering that the viscosity in the radiative zone should be larger by a factor of at least $10^{4}$ than the molecular value to account for the rigid  rotation of the solar radiative zone \citep[e.g.,][]{RuedigerKichatinov96,Spadaetal10}, such a lengthscale may increase up to $10^{6}$~m, even without invoking the turbulence produced by the oscillations of the resonant flux tube once these reach a sizeble amplitude. If the magnetic field in the overshoot layer has a spatially intermittent structure with the resonant flux tube close to the strong-field flux tubes with a mean separation comparable with or smaller than $\lambda_{\rm F}^{-1}$, the ensemble of active regions formed in the photosphere will reproduce the same pattern  of the autoresonant oscillation mode. Specifically, different strong-field flux tubes may become unstable with their $m=1$ or $m=2$ growing oscillations in spatial coincidence with the nodes of the oscillations of the autoresonant flux tube. 
The pattern of photospheric active regions will statistically display the same azimuthal wavenumber of the oscillating flux tube. The longitude of the starspots observed in the hot Jupiter host Kepler-17 may be accounted for by such a mechanism \citep[cf. Sect.~2 of][for a description of the observations]{Lanza22}. 

The  potential magnetic field in the stellar corona corresponding to a photospheric pattern dominated by a mode with azimuthal order $m$ will have a degree $l \geq m$. Since $m$ can be as large as 8 \citep[cf.][]{Lanza22}, the corresponding increase in the degree $l$ of the coronal field will strongly reduce the Alfven radius of the stellar wind and consequently the angular momentum loss rate \citep{Revilleetal15}. Specifically, considering the fitting formula provided by \citet{Revilleetal15} and based on numerical simulations, the torque for given stellar radius, angular velocity, and mass loss rate scales with an exponent approximately of $2/(2l+2.7)$. For solar-like parameters, their models find a decrease by a factor of $\sim 4$ in passing from the  dipolar ($l=1$) to the quadrupolar ($l=2$) configuration. Going from the dipolar to the configuration with $l=8$, their analytic formula gives a reduction by a factor $\sim 32$. Considering that the actual coronal field configuration consists of the contributions of components with different values of $l$, we shall prudently adopt a reduction by a factor of 10 of the stellar wind torque when autoresonant oscillations are excited. The corresponding slower evolution of the stellar spin allows a system to enter and remain locked into an autoresonant regime  for long time intervals as already envisaged in Sect.~3.9.2 of \citet{Lanza22} and as we shall see in Sect.~\ref{evolution_of_omega_mn}. 

{Finally, we note that the non-linear development of the instability of the strong-field flux tube  produces a perturbation on the resonant tube itself having a duration significantly shorter than the period of its oscillations because the unstable flux tube emerge to the photosphere in approximately one month \citep{Caligarietal95}. In other words, the perturbation effects are averaged to zero on the timescales characteristic of the amplification of the oscillations and thus have virtually no effect on the resonant flux tube and the autoresonant process (cf. Sects.~\ref{autoresonance_section} and~\ref{applications}). }

\subsection{Tidal dissipation and torque in autoresonant systems}
\label{auto_tidal_tor}
The equations governing the oscillations of the autoresonant flux tubes are Eqs.~\eqref{eq_motion1} and~\eqref{eq_motion2}. Following the method introduced in Sect.~3.4 of \citet{Lanza22}, it is possible to write their first integral, that is, the equation of conservation of the mechanical energy, in the form
\begin{equation}
    \frac{d{\cal E}}{dt} = -2\zeta \rho (\dot{\xi}_{\phi}^{2} + \dot{\xi}_{r}^{2}) - \rho \dot{\vec \xi} \cdot \nabla \Psi_{lmn},
\end{equation}
where $\cal E$, the total mechanical energy density of the oscillating flux tube, consists of the kinetic energy, the magnetic tension energy, and the potential energy coming from the work done against the stable stratification of the overshoot layer by the radial displacement of the oscillations \citep[see][for details]{Lanza22}. 

The power dissipated per unit volume is given by the dissipation function
\begin{equation}
    p_{\rm diss} = -2 \zeta \rho (\dot{\xi}_{\phi}^{2} + \dot{\xi}_{r}^{2} ) \simeq - 2 \zeta \rho \dot{\xi}_{\phi}^{2}
\end{equation}
because $\xi_{r} \ll \xi_{\phi}$ in resonance. The total dissipated power is obtained by integrating $p_{\rm diss}$ over the volume of the oscillating flux tube. Since its density can be regarded as uniform and its volume is $V_{\rm ft} = (1/2) \pi^{2} r_{0} d_{\rm ft}^{2}$, where $d_{\rm ft}$ is the diameter of the cross section of the flux tube, we find the total dissipated power, averaged over one period of the oscillations,
\begin{equation}
{\cal P}_{\rm diss} = - \frac{1}{2} \pi^{2} r_{0} d_{\rm ft}^{2} \zeta \rho \omega_{mn}^{2} \hat{\xi}^{2}_{\phi},
\label{diss_power}
\end{equation}
where we substituted $\dot{\xi}_{\phi} = \omega_{mn} \xi_{\phi}$ and considered that the average of $\xi_{\phi}^{2}$ over one oscillation period is $\hat{\xi}_{\phi}^{2}/2$. 

Equation~\eqref{diss_power} gives the dissipated power in the reference frame rotating with the star due to the oscillations excited by the tidal potential \citep[cf. Sect.~2.2 of][]{Ogilvie14}. The corresponding tidal torque, ${\cal T}$, acting on the stellar spin is given by \citep[cf. Sect.~2.2 of][]{Ogilvie14}
\begin{equation}
    {\cal T} = \frac{m}{\omega_{mn}}{\cal P}_{\rm diss} = \frac{1}{2} \pi^{2} m r_{0} d_{\rm ft}^{2} \zeta \rho \omega_{mn} \, \hat{\xi}_{\phi}^{2}.
    \label{tidal_torque_eq}
\end{equation}
 The maximum tidal torque corresponds to the maximum amplitude of the oscillations  determined according to the method illustrated in the final paragraph of Sect.~\ref{autoresonance_section}, that is, ${\cal T}_{\max} = {\cal T}(\hat{\xi}_{\phi \max})$. The corresponding timescale for the variation in the stellar angular velocity is
 \begin{equation}
     \tau_{\rm tide} = \frac{I_{\rm s} \Omega_{\rm s}}{{\cal T}_{\max}}. 
     \label{tidal_time_autores}
 \end{equation}
 
 \subsection{Evolution of the tidal potential frequency $\omega_{mn}$}
 \label{evolution_of_omega_mn}
In principle, the evolution of the tidal potential frequency $\omega_{mn}$ as given by Eq.~\eqref{omega_mn_dot} can be computed if the effects of the stellar wind and the tides on the stellar spin and the orbital mean motion can be evaluated. These require the knowledge of the angular momentum loss rate and the modified stellar tidal quality factor $Q^{\prime}_{\rm s}$ together with a numerical integration in time of the differential equations~\eqref{star_rot_evol_eq} and~\eqref{mean_motion_var} for each particular system. Such a detailed investigation is beyond the scope of the present work that focuses on the excitation of the oscillations in the resonant magnetic flux tubes and will be postponed to future analyses. Nevertheless, some consideration of the evolution of the tidal  frequency $\omega_{mn}$ is in order here, especially in view of a preliminary estimation of the duration of the autoresonant phase in the observed systems. For this reason, a theory for the evolution of $\omega_{mn}$ is developed in this section. 
 
 To apply the results of the previous sections to the evolution of the tidal potential frequency $\omega_{mn}$, we write the equation giving the evolution of the orbital angular momentum as 
 \begin{equation}
     \frac{dL_{\rm orb}}{dt} = - K (\Omega_{0} - \Omega_{\rm s})
     \label{time_ev_L_orb}
 \end{equation}
 and recast Eqs.~\eqref{star_rot_evol_eq} as 
 \begin{equation}
     I_{\rm s} \frac{d\Omega_{\rm s}}{dt} = K (\Omega_{0} - \Omega_{\rm s}) + \frac{dL_{\rm w}}{dt},
 \end{equation}
where 
\begin{equation}
    K \equiv \frac{9}{2 r_{\rm g}^{2} Q^{\prime}_{\rm s}} \left( \frac{m_{\rm p}}{M_{\rm s}} \right)^{2} \left( \frac{R_{\rm s}}{a}\right)^{3} I_{\rm s} \Omega_{0},
\end{equation}
that is valid for $m_{\rm p} \ll M_{\rm s}$ with $r_{\rm g}^{2}= 0.07$ being the square of the nondimensional gyration radius of the host star. From Kepler third law and Eq.~\eqref{time_ev_L_orb}, we have 
\begin{equation}
    \frac{d\Omega_{0}}{dt} = \frac{3K}{m_{\rm red} a^{2}} (\Omega_{0} - \Omega_{\rm s}),
\end{equation}
where $m_{\rm red} = M_{\rm s}m_{\rm p}/(M_{\rm s}+m_{\rm p}) \sim m_{\rm p}$ is the reduced mass of the system. In such a way, Eq.~\eqref{omega_mn_dot} can be recast as
\begin{equation}
    \frac{d\omega_{mn}}{dt} = \frac{n}{I_{\rm s}} \, K {\cal H}(m,n) \, (\Omega_{0} - \Omega_{\rm s}) + \frac{m}{I_{\rm s}} \frac{dL_{\rm w}}{dt}, 
    \label{omega_dot_with_h}
\end{equation}
where 
\begin{equation}
    {\cal H}(m,n) \equiv \left(\frac{m}{n}\right) - \left(\frac{3I_{\rm s}}{m_{\rm red} a^{2}}\right). 
    \label{hmn_defin}
\end{equation}
In the systems were tides are negligible, $\Omega_{0}$ is constant and we have 
\begin{equation}
\frac{d\omega_{mn}}{dt} = \left( \frac{m}{I_{\rm s}} \right) \frac{dL_{\rm w}}{dt} = m \frac{d\Omega_{\rm s}}{dt} 
\end{equation}
and the possibility of having $|\kappa| $ smaller than the threshold for the excitation of autoresonant oscillations is linked to a phase of very slow variation in the stellar spin, in other words, a phase of stalled rotational evolution. Once autoresonant oscillations are triggered and reach a sizeable amplitude, they can slow down the rotational evolution by reducing the coronal Alfven radius as proposed in Sect.~\ref{auto_wind_tor}. This contributes to increase the duration of the phase of stalled rotation  because the reduced loss rate of angular momentum from the convection zone makes the reservoir of angular momentum in the radiative zone sufficient to balance the losses for a longer time. By decreasing $dL_{\rm w}/dt$ by a factor of, say, ten, the duration of the phase of stalled rotation may be increased by a corresponding factor reaching $1-10$~Gyr according to the mass of the host star, that is, comparable with the main-sequence lifetime of the system. Therefore, we expect a long duration of the commensurability phase in such a regime, up to several Gyr, especially for lower main-sequence hosts for which the stalled rotation phase is longer.  

On the other hand, in systems where the tidal torque is non-negligible,  the angular momentum loss in the wind may be balanced by the torque itself making $d\omega_{mn}/dt \sim 0$ and $|\kappa|$ smaller than the threshold for triggering autoresonance. 
Considering that $dL_{\rm w}/dt < 0$ and $K/I_{\rm s} >0$ in Eq.~\eqref{omega_dot_with_h}, when $\Omega_{0} > \Omega_{\rm s}$ this implies  ${\cal H}(m,n) > 0$, or  ${\cal H} (m,n) < 0$ when $\Omega_{0} < \Omega_{\rm s}$. We shall see examples of star-planet systems verifying both kinds of prescriptions in Sect.~\ref{non-synchronous_systems}.

By writing $\omega_{mn} = n(\Omega_{\rm s}- \Omega_{0}) + p \Omega_{\rm }$, where $p \equiv m-n$ is a positive or negative integer, we can recast Eq.~\eqref{omega_dot_with_h} as
\begin{equation}
    \frac{d\omega_{mn}}{dt} = -\Theta\, \omega_{mn} + \left(p \Theta + \frac{m}{L_{\rm s}}\frac{dL_{\rm w}}{dt} \right) \Omega_{\rm s},
    \label{omega_dot_ode}
\end{equation}
where  
\begin{equation}
    \Theta \equiv \frac{K}{I_{\rm s}} {\cal H}(m,n) 
    \label{definition_Hmn}
\end{equation}
with $|\Theta| $ being the inverse timescale for the evolution of $\omega_{mn}$ and $L_{\rm s} \equiv I_{\rm s} \Omega_{\rm s}$ the angular momentum of the stellar spin. 

The general solution of the non-homogeneous differential equation~\eqref{omega_dot_ode} can be found by the method of the variation of the constant and is
\begin{eqnarray}
    \omega_{mn} (t)  =  C_{0} \exp (-\Theta t) + \nonumber \\
     \exp (-\Theta t) \int \left(p \Theta + \frac{m}{L_{\rm s}(t)}\frac{dL_{\rm w}}{dt} \right) \Omega_{\rm s} (t) \exp(\Theta t) \, dt, 
     \label{omega_ode_sol}
\end{eqnarray}
where $C_{0}$ is a constant that is determined by the initial condition. 

The frequency of the autoresonant magnetostrophic waves considered in our model is much smaller than the stellar spin frequency ($|\omega_{mn}| \ll \Omega_{\rm s}$), therefore, the initial condition can be well approximated as $\omega_{mn} (0) \simeq 0$ that allows us to find $C_{0}$. The indefinite integral appearing in Eq.~\eqref{omega_ode_sol} can be evaluated only when the evolution of the stellar spin, of the tidal torque, and the angular momentum loss rate are known. Nevertheless, if the evolution of $\Omega_{\rm s}$ is very slow because of the balance between the tidal torque and the angular momentum loss rate of the wind (with $\Omega_{0} > \Omega_{\rm s}$) and $|dL_{\rm w}/dt|$ is made small because of the reduced Alfven radius in the autoresonant regime (cf. Sect.~\ref{auto_wind_tor}) or the presence of a massive close-by planet (cf. Sect.~\ref{frequency_var_general}), the integrand in Eq.~\eqref{omega_ode_sol} can be regarded as approximately constant and the solution becomes
\begin{equation}
    \omega_{mn} (t) \simeq \left(p - \frac{m}{\Theta} \frac{1}{\tau_{\rm spin}} \right) \Omega_{\rm s} \left[1- \exp(-\Theta t) \right],
    \label{omega_ode_const}
\end{equation}
where $\tau_{\rm spin} \equiv -L_{\rm s}/(dL_{\rm w}/dt)> 0 $ is the timescale for the evolution of the stellar spin under the action of the reduced angular momentum loss in the stellar wind without considering the action of tides. 

Restricting our analysis to systems with $\Omega_{0} > \Omega_{\rm s}$ and assuming that the tidal torque is balancing the angular momentum loss in the wind, we have ${\cal H}(m,n) > 0$ that implies $\Theta > 0$ according to Eq.~\eqref{definition_Hmn}.  When the system is close to  commensurability, $|\omega_{mn}| \ll \Omega_{\rm s}$ and  the second term dominates in the right-hand side of Eq.~\eqref{omega_dot_ode}. In such a case, the condition for the system to enter the autoresonant regime as given by Eq.~\eqref{kappa_cond} becomes
\begin{equation}
   | \kappa |= \left|  -\frac{1}{\omega_{mn}^{2}} \frac{d\omega_{mn}}{dt} \right| \simeq \left| p\Theta - \frac{m}{\tau_{\rm spin}} \right| \frac{\Omega_{\rm s}}{\omega_{mn}^{2}} \leq \kappa_{\rm th},
\end{equation}
that is satisfied for $p >0$ provided that 
\begin{equation}
 \tau_{\rm spin \min} \leq  \tau_{\rm spin} \leq \tau_{\rm spin \max}
    \label{tau_spin_cond}
\end{equation}
with
\begin{equation}
      \tau_{\rm spin \min} \equiv \frac{m}{p\Theta + \kappa_{\rm th} \omega_{mn}^{2}/\Omega_{\rm s}} 
      \label{tau_spin_min}
\end{equation} and
\begin{equation}
\tau_{\rm spin \max} = \frac{m}{p\Theta - \kappa_{\rm th} \omega_{mn}^{2}/\Omega_{\rm s}}. 
\label{tau_spin_max}
\end{equation}
In some systems, $ p \Theta \leq \kappa_{\rm th} \omega_{mn}^{2}/\Omega_{\rm s}$. In those cases, $\tau_{\rm spin \max}$ is not defined and we have the only condition $\tau_{\rm spin} \geq \tau_{\rm spin \min}$. 

The asymptotic value of $\omega_{mn}$, that is, its value when $ t \gg \Theta^{-1}$, is given by $p-m/(\Theta \tau_{\rm spin})$ that is very small if $\tau_{\rm spin}$ is close to $m /(p \Theta)$ which falls in the mid of the interval of the allowed values as given by Eq.~\eqref{tau_spin_cond} or, in any case, is larger than $\tau_{\rm spin \min}$. In such a case, $\omega_{mn}$ remains close to its initial zero value all along the system evolution. In other words, if the host star has $\tau_{\rm spin} \sim m/(p\Theta)$, the system  remains close to commensurability for a very long time interval comparable with its main-sequence lifetime. This is more likely to occur if $p\Theta / \omega_{mn} \gg \kappa_{\rm th} (\omega_{mn}/\Omega_{\rm s})$. 

When  condition~\eqref{tau_spin_cond} is violated in a system with $\Omega_{0} > \Omega_{\rm s}$, for example, if $\tau_{\rm spin} > \tau_{\rm spin \max}$, the wind braking will become too weak and the system may exit the autoresonant regime because of the tidal effects. As a consequence,  the angular momentum loss rate by the wind will increase because a lower $l$ coronal field will become dominant  (cf. Sect.~\ref{auto_wind_tor}). This will restore  condition~\eqref{tau_spin_cond} on $ \tau_{\rm spin}$ bringing the system back into autoresonance. Similarly, if for some reason $\tau_{\rm spin} < \tau_{\rm spin \min}$,  the system will exit the autoresonant regime, the wind braking will strongly increase and the stellar rotation will be slowed down increasing the difference $\Omega_{0} - \Omega_{\rm s}$, that is, the tidal torque that tends to bring back the system into autoresonance. 

We assume that the value of $\tau_{\rm spin}$ may be adjusted in order to verify condition~\eqref{tau_spin_cond} because it depends on the amplitude of the autoresonant oscillations that are responsible for the perturbation of the unstable flux tubes in the overshoot layer (cf. Sect.~\ref{auto_wind_tor}). When the oscillation amplitude increases, the perturbation will be greater making the coronal field more strongly dominated by the high-$l$ mode that makes the wind braking less efficient. In such a way, the value of $\tau_{\rm spin}$ can be increased or decreased according to the amplitude of the excited oscillations. 

In conclusion, we foresee  a self-regulating mechanism that can keep systems with $\Omega_{0} > \Omega_{\rm s}$ and $m> n$ in autoresonance with comparable tidal and wind torques  for long time intervals, comparable with the main-sequence lifetimes of their host stars.  

\section{Applications}
\label{applications}
The theory introduced in Sect.~\ref{model} will now be applied to ten star-planet systems showing a commensurability between the star spin period and the orbital period of their closest planet already investigated by \citet{Lanza22}. The purpose of that work was to account for the observed $P_{\rm orb}/P_{\rm rot} = m/n$ commensurability as the result of the tidal excitation of magnetostrophic oscillations  inside a toroidal flux tube stored in the overshoot layer of the host star. The results of the present investigation  allow us to study in a quantitative way the  excitation of those oscillations and the maintainance of each system in autoresonance for a sufficiently long time interval  as explained in Sect.~\ref{model}. 

The relevant parameters of the considered systems are listed in Table~\ref{table1param} that is extracted from Table~1 of \citet{Lanza22} with an update of the mass of AU~Mic~b. In Table~\ref{table1param}, we list from the left to the right, the name of the system, the mass, $M_{\rm s}$, and  the radius, $R_{\rm s}$, of the host star, its effective temperature, $T_{\rm eff}$, the orbit semimajor axis, $a$, and its ratio to the host radius, $a/R_{\rm s}$, the eccentricity of the orbit, $e$, the mass of the planet, $m_{\rm p}$, the sky-projected spin orbit angle, $\lambda_{\rm so}$, the star rotation period, $P_{\rm rot}$, the orbital period, $P_{\rm orb}$, the integers $m$ and $n$ the ratio of which best approximate the $P_{\rm rot}/P_{\rm orb}$ ratio, and the relevant references. 

Following the same approach as in \citet{Lanza22}, for the sake of simplicity we neglect the (small) eccentricity of the planetary orbits in the computation of the components $\Psi_{lmn}$ of the tidal potential and keep only the effects of the obliquity $i$ that we assume to be $20^{\circ}$ for all the systems, except for Kepler-13A ($i=25^{\circ}$), Kepler-63 ($i=-120^{\circ}$), HAT-P-11 ($i=70^{\circ}$), and WASP-107 ($i=40^{\circ}$). Their specific obliquity values are derived from the measurements of their sky-projected spin-orbit obliquity $\lambda_{\rm so}$ in Table~\ref{table1param} together with an estimate of the inclination of their spin axis to the line of sight based on the stellar rotation period and the inclination of the orbital plane given by the transit best fit. The value of the degree $l$ of the tidal potential component is constrained to be such that $l \geq |m|$ and $l-n$ be even, otherwise the corresponding $\Psi_{lmn}$ component has a zero $A_{lmn}(0,i)$ coefficient and does not contribute to the excitation of the oscillations \citep[cf.][]{Ogilvie14}. In Table~\ref{tablelmn}, we list, from the left to the right, the name of the system, the fractionary radius at the base of the convection zone, $r_{0}/R_{\rm s}$, the density in that layer, $\rho$, the minimum $l$ compatible with these constraints together with the corresponding values of $m$, $n$, and the absolute value of the tidal coefficient, $A_{lmn}$, computed as in Sect.~3.1 of \citet{Lanza22}. 
\begin{table*}
\begin{small}
\caption{Systems showing evidence of spin-orbit commensurability \citep[cf.][Table~1]{Lanza22}.} 
\begin{center}
\begin{tabular}{ccccccccccccc}
\hline
System & $M/M_{\odot}$ & $R/R_{\odot}$ &  $T_{\rm eff}$ & $a$  & $a/R$ & $e$ & $m_{\rm p}$ & $\lambda_{\rm so}$ & $P_{\rm rot}$ & $P_{\rm orb}$ & $P_{\rm rot}$:$P_{\rm orb}$  & References \\
& ($M_{\odot}$) & ($R_{\odot}$) & (K) & (au) & & & ($M_{\rm J}$) & (deg) & (day) & (day) & ($m$:$n$) &   \\
\hline
\object{CoRoT-2} & 0.97 & 0.902 & 5625 & 0.0281 & 6.70 & 0.0 &  3.30 & 0.0 & 4.552 & 1.743 & 8:3 & 1, 2 \\
\object{CoRoT-4} & 1.16 & 1.17 & 6190 & 0.0902 & 17.36 & $<0.14$ & 0.703 & & 9.202 & 9.202 & 1:1 & 1, 3\\ 
\object{CoRoT-6} & 1.05 & 1.025 & 6090 & 0.0854 & 17.94 & $< 0.18$ & 2.95 & & 6.35 & 8.887 & 5:7 & 1, 4\\
\object{Kepler-13A} & 1.72 & 1.74 & 7650 & 0.0355 & 4.44  &  & $4.9-8.1$ & $23\pm 4$ & 1.046 & 1.764 & 3:5 & 5, 6, 7\\
\object{Kepler-17} & 1.16 & 1.05 & 5780 & 0.0268 & 5.48 & $ <0.02$ & 2.47 & $15\pm 15$ & 11.89 & 1.486 & 8:1 & 1, 8, 9\\
\object{Kepler-63} & 0.984 & 0.90 & 5575 & 0.080 & 19.1 & $<0.45$ & $ <0.38$ & $-110\pm 20$  & 5.401 & 9.424 & 3:5 & 10\\
\object{HAT-P-11} & 0.81 & 0.75 & 4780 & 0.053 & 15.6 & 0.218 & 0.0736 & $90\pm 30$ & 30.5 & 4.888 & 6:1 & 11, 12, 13, 14\\
\object{$\tau$ Boo} & 1.39 & 1.42 & 6400 & 0.049 & 7.40 & $< 0.011$ & 6.13 &  & 3.31 & 3.3124 & 1:1 & 15, 16 , 17\\
\object{WASP-107} & 0.69 & 0.66 & 4430 & 0.0558 & 18.16 &  0.0 & 0.12 & $40-140$ & 17.1 & 5.721 & 3:1 & 18 \\
\object{AU Mic} & 0.50 & 0.75 & 3700 & 0.0678 & 19.2 & 0.181 & 0.036 & $-3 \pm 10$  & 4.8367 & 8.463 & 4:7 & 19, 20, 21 \\
\hline
\label{table1param}
\end{tabular}
\end{center}
Note. References: 1: \citet{Bonomoetal17}; 2: \citet{Lanzaetal09_C2}; 3: \citet{Lanzaetal09}; 4: \citet{Lanzaetal11}; 5: \citet{Szaboetal14}; 6: \citet{Shporeretal14}; 7: \citet{Estevesetal15}; 8: \citet{Desertetal11}; 9: \citet{Lanzaetal19}; 10: \citet{Sanchis-Ojedaetal13}; 11: \citet{Bakosetal10}; 12: \citet{Sanchis-Ojedaetal11}; 13: \citet{Bekyetal14}; 14: \citet{Yeeetal18}; 15: \citet{Walkeretal08}; 16: \citet{Brogietal12}; 17: \citet{Borsaetal15}; 18: \citet{DaiWinn17}; 19: \citet{Szaboetal21}; 20: \citet{Caleetal21}; 21: \citet{Zicheretal22}. 
\end{small}
\label{table1}
\end{table*}
\begin{table}
\caption{Structure and tidal potential parameters.}
\begin{center}
\begin{tabular}{ccccccc}       
\hline                
                                                          System & $r_{0}/R_{\rm s}$ & $\rho$ & $l$ & $m$ & $n$ & $|A_{lmn}(0,i)|$ \\
             & & (kg m$^{-3}$) & & & & \\
             \hline 
        CoRoT-2 &      0.726 &       3.29e+02 &     9 &     8 &     3 &   5.0489e-03 \\
        CoRoT-4 &      0.805 &       2.26e+01 &     3 &     1 &     1 &   4.0068e-01 \\
        CoRoT-6 &      0.744 &       1.86e+02 &     7 &     5 &     7 &   1.0028e-01 \\
     Kepler-13A &      0.120 &       3.31e+04 &     5 &     3 &     5 &   1.3753e-01 \\
      Kepler-17 &      0.795 &       4.66e+01 &     9 &     8 &     1 &   3.2176e-04 \\
      Kepler-63 &      0.726 &       3.29e+02 &     5 &     3 &     5 &   1.0420e-02 \\
       HAT-P-11 &      0.675 &       7.78e+02 &     7 &     6 &     1 &   8.3935e-02 \\
    $\tau$~Boo &      0.868 &       8.65e-01 &     3 &     1 &     1 &   4.0068e-01 \\
       WASP-107 &      0.673 &       1.79e+03 &     3 &     3 &     1 &   2.0495e-01 \\
          AU~Mic &      0.303 &       6.36e+03 &     7 &     4 &     7 &   3.5363e-02 \\
\hline
\end{tabular}
\label{tablelmn}
\end{center}
\end{table}
\begin{table*}
\caption{Parameters of magnetic flux tube autoresonant oscillations.}
\begin{center}
\begin{tabular}{cccccccccccc}       
\hline  
      System & $B_{\rm eq}$ & $\beta$ & $T$ & $\tau$ & $\omega_{0}$ & $\omega_{0}/\Omega_{\rm s}$ & $\gamma_{\rm D}$ & $\kappa_{\rm th}$ & $\kappa_{\rm exp}$ & $8\kappa_{\rm th}\omega_{0}/(3 |\gamma_{\rm D}|)$ \\
                        & (T) & & & (s) & (s$^{-1}$) & & (m$^{-2}$) & & & (m$^{2}$ yr$^{-1}$) \\
\hline
        CoRoT-2 &      0.203 &     6.44e+08 &    -4.51e+03 &     7.12e+06 &        4.88e-08 &        3.05e-03 &       -1.19e-17 &        1.53e-07 &        6.51e-06 &        5.29e+10 \\
        CoRoT-4 &      0.053 &     3.46e+08 &    -1.53e+03 &     7.38e+06 &        4.55e-09 &        5.81e-04 &       -1.03e-19 &        5.95e-01 &        5.66e-06 &        2.20e+18 \\
        CoRoT-6 &      0.153 &     5.73e+08 &    -2.84e+03 &     7.17e+06 &        3.03e-08 &        2.64e-03 &       -4.59e-18 &        8.10e-09 &        3.29e-06 &        4.50e+09 \\
     Kepler-13A &      2.039 &     3.53e+09 &    -2.75e+05 &     1.52e+07 &        4.83e-08 &        6.95e-04 &       -1.17e-17 &        8.84e-05 &        2.10e-05 &        3.08e+13 \\
      Kepler-17 &      0.077 &     4.11e+08 &    -1.27e+03 &     6.85e+06 &        5.05e-08 &        8.26e-03 &       -1.27e-17 &        9.88e-08 &        1.73e-07 &        3.29e+10 \\
      Kepler-63 &      0.203 &     6.44e+08 &    -3.59e+03 &     7.12e+06 &        1.92e-08 &        1.43e-03 &       -1.85e-18 &        3.17e-07 &        9.44e-06 &        2.78e+11 \\
       HAT-P-11 &      0.313 &     7.33e+08 &    -1.55e+03 &     7.14e+06 &        1.20e-07 &        5.04e-02 &       -7.22e-17 &        8.80e-13 &        4.49e-09 &        1.23e+05 \\
         $\tau$~Boo &      0.010 &     1.71e+08 &    -2.85e+03 &     4.66e+06 &        2.93e-09 &        1.35e-04 &       -4.30e-20 &        8.95e+02 &        1.74e-04 &        5.14e+21 \\
       WASP-107 &      0.474 &     7.72e+08 &    -1.73e+03 &     6.07e+06 &        6.04e-08 &        1.42e-02 &       -1.83e-17 &        8.23e-03 &        8.24e-08 &        2.29e+15 \\
          AU~Mic &      0.894 &     1.07e+09 &    -3.00e+04 &     2.25e+07 &        5.93e-08 &        3.94e-03 &       -1.76e-17 &        9.19e-15 &        5.55e-06 &        2.61e+03 \\
\hline                                                                    
\end{tabular}
\label{table_auto_res_param}
\end{center}
\end{table*}

To apply our theory, we assume that the magnetic field of the resonant  flux tube in the overshoot layer is in equipartition with a plasma velocity field of amplitude $\varv_{\rm d} = 10$~m~s$^{-1}$.  The corresponding field intensity is $B_{\rm eq} = \varv_{\rm d}\sqrt{\mu \rho}$. This ensure that the flux tube can neither be disrupted nor significantly distorted in the lower part of the overshoot layer where we assume that the velocity of the downdrafts has been braked well below $10$~m~s$^{-1}$. 

The equilibrium azimuthal velocity $\varv_{0}$ inside the resonant flux tube is assumed to be equal to the Alfven velocity (cf. Sect.~\ref{oscill_freq_sect}), that is,  the maximum value compatible with stability against the Kelvin-Helmoltz instability \citep{Ferriz-MasSchussler93}. However, the effect of the internal velocity $\varv_{0}$ on the flux tube parameters is very small because it is much lower than the rotation velocity, that is, $\varv_{0} = \varv_{\rm A} \ll \Omega_{\rm s} r_{0}$.

The parameters governing the flux tube oscillations are reported in Table~\ref{table_auto_res_param} where we list, from the left to the right, the name of the system, the equipartition magnetic field, $B_{\rm eq}$, of the oscillating flux tube, the parameter $\beta \equiv 2\mu p/B_{\rm eq}^{2}$, the non-dimensional stratification parameter, $T$, as given by Eq.~(8) of \citet{Ferriz-MasSchussler94} (its negative values indicate a stable stratification of the overshoot layer, that is, $\delta < 0$), the characteristic timescale $\tau = \sqrt{2}H/\varv_{\rm A}$ with $H$ being the pressure scale height at $r=r_{0}$ \citep[see Sect.~3 of][]{Ferriz-MasSchussler94}, the frequency, $\omega_{0}$, of the magnetostrophic oscillations, its ratio to the spin frequency, $\omega_{0}/\Omega_{\rm s}$, the coefficient $\gamma_{\rm D}$ as appearing in our  Duffing oscillator equation~\eqref{duffing_osc}, the threshold frequency drift, $\kappa_{\rm th}$, as given by Eq.~\eqref{kappa_thres}, the frequency drift expected on the basis of the present parameters of the system, $\kappa_{\rm exp}$, as computed from Eqs.~\eqref{kappa_vs_omega_mn}, \eqref{omega_mn_dot}, \eqref{star_rot_evol_eq}, \eqref{angular_mom_loss_rate}, and~\eqref{mean_motion_var}, and the parameter giving the asymptotic amplitude $A$ vs. the time at the threshold drift as $A \sim \{[8\kappa_{\rm th}\omega_{0}/(3|\gamma_{\rm D}|)] t \}^{1/2}$.  To compute the parameters listed in Table~\ref{table_auto_res_param}, we made use of the parameters listed in Table~3 of \citet{Lanza22} and assumed a superadiabaticity $\delta =-2.0 \times 10^{-6}$, an adiabatic exponent $\gamma = 5/3$, and a differential rotation parameter $q=0.06$ \citep[cf.][]{Ferriz-MasSchussler94}. 
\begin{table}[]
    \centering
        \caption{Coefficients of transformation from $(\hat{\xi}_{\phi}, \hat{\xi}_{r})$ to the normal oscillation modes $\hat{\Theta}_{1}, \hat{\Theta}_{2}$ (cf. Eqs.~\ref{norm_mode1} and~\ref{norm_mode2}). }
    \begin{tabular}{ccc}
                         \hline
                                     System & $(\Lambda_{1}-C)/D$ & $(\Lambda_{2}-C)/D$ \\
   \hline
         CoRoT-2     &      -0.0177     &      56.52 \\
         CoRoT-4     &      -0.0026     &     388.63 \\
         CoRoT-6     &      -0.0127     &      78.76 \\
      Kepler-13A     &      -0.0056     &     178.67 \\
       Kepler-17     &      -0.0232     &      43.18 \\
       Kepler-63     &      -0.0074     &     135.29 \\
        HAT-P-11     &      -0.0191     &      52.28 \\
          $\tau$ Boo     &      -0.0010     &    1011.38 \\
        WASP-107     &      -0.0111     &      90.18 \\
           AU~Mic     &      -0.0296     &      33.83 \\
\hline
    \end{tabular}
    \label{normal_mode_table}
\end{table}

\begin{table*}
\caption{Parameters ruling the evolution of the tidal frequency $\omega_{mn}$.}
\begin{center}
\begin{tabular}{ccccccccccccc}       
\hline  
Systems  & ${\cal H}(m,n)$ & $Q_{\rm s \ eq}^{\prime}$  & $\hat{\xi}_{\phi \max}$ &   $\tau_{\rm tide}$ & $\tau_{\rm red \ AMLR}$ & $\Theta^{-1}$ & $\tau_{\rm spin \min}$ & $\tau_{\rm spin \max}$ & $\tau_{\rm spin \ ne}$ \\
                                                                                          & & & (m) & (yr) & (yr) & (yr) & (yr) & (yr) & (yr) \\
\hline
              CoRoT-2 &        1.214 &      1.09e+07 &      8.16e+06 &      3.80e+16 &      2.60e+09 &      3.34e+09 &      3.61e+09 &      1.03e+10 &      5.35e+09 \\
              CoRoT-4 &       -0.403 &      &      3.11e+08 &      2.26e+16 &      2.12e+10 &     -3.73e+13 &      2.02e+04 &      &      \\
              CoRoT-6 &        0.504 &      &      1.13e+06 &      7.03e+18 &      6.01e+09 &      2.39e+12 &      2.55e+11 &     &     \\
           Kepler-13A &       -3.613 &   &      2.93e+07 &      7.12e+15 &      1.18e+09 &     -6.76e+08 &      3.11e+07 &      &      1.01e+09 \\
            Kepler-17 &        4.391 &   6.08e+08   &      1.60e+06 &      3.28e+18 &      2.93e+10 &      1.51e+10 &      4.53e+09 &      &      1.72e+10 \\
            Kepler-63 &       -0.952 &   &      4.21e+07 &      8.13e+15 &      3.67e+09 &     -7.34e+13 &      1.09e+10 &      &      1.10e+14 \\
             HAT-P-11 &       -5.505 &      &      1.39e+02 &      3.43e+24 &      7.00e+10 &     -4.26e+14 &      3.84e+13 &      &      \\
               $\tau$~Boo &        0.108 &       &      4.82e+08 &      1.38e+18 &      4.60e+09 &      1.01e+11 &      8.96e+01 &    &   \\
             WASP-107 &       -0.771 &      &      2.34e+07 &      2.68e+14 &      1.34e+10 &     -2.15e+15 &      1.34e+04 &     &    \\
                AU~Mic &       -8.411 &   &      2.67e+02 &      3.03e+24 &      9.77e+08 &     -1.81e+14 &      2.40e+14 &      2.42e+14 &      2.41e+14 \\
\hline                                                                    
\end{tabular}
\label{table_tidal_freq_param}
\end{center}
\end{table*}

{As a preliminary step to the application of our model, we check that the normal modes $\hat{\Theta}_{1}$ and $\hat{\Theta}_{2}$ verify the approximations in Eqs.~\eqref{norm_mode1} and~\eqref{norm_mode2} by listing in Table~\ref{normal_mode_table} the values of the ratios $(\Lambda_i-C)/D$, with $i=1,2$. We see that $\hat{\Theta}_{2} \simeq \hat{\xi}_{\phi}$ is an acceptable approximation for all our investigated systems. }

The parameters describing the evolution of the frequency of the tidal potential $\omega_{mn}$ in resonance with the flux tube magnetostrophic frequency $\omega_{0}$ are listed in Table~\ref{table_tidal_freq_param} where we report, from the left to the right, the name of the system, ${\cal H}(m,n)$, as defined by Eq.~\eqref{hmn_defin}, the modified tidal quality factor, $Q^{\prime}_{\rm s \ eq}$, corresponding to the equilibrium between the tidal torque and the wind braking torque in the non-synchronous systems where $\Omega_{0} > \Omega_{\rm s}$ and ${\cal H}(m,n) > 0$, the maximum amplitude of the azimuthal oscillations, $\hat{\xi}_{\phi \max}$, estimated with the method in the final paragraph of Sect.~\ref{autoresonance_section}, the tidal timescale, $\tau_{\rm tide}$, associated with the tidal torque due to the autoresonant oscillations as given by Eq.~\eqref{tidal_time_autores} and $2\zeta=1.5 \times 10^{-15}$~s$^{-1}$, the braking timescale of the stellar wind, $\tau_{\rm red \ AMLR}$, in the case of a magnetic braking reduced by a factor of ten with respect to Eq.~\eqref{angular_mom_loss_rate}, the timescale $\Theta^{-1}$ as derived from Eq.~\eqref{definition_Hmn}, the minimum and maximum spin timescales, $\tau_{\rm spin \min}$, and $\tau_{\rm spin \max}$, as derived from Eqs.~\eqref{tau_spin_min} and~\ref{tau_spin_max}, respectively, and the spin timescale $\tau_{\rm spin \ ne} = m/(p\Theta)$ giving $d\omega_{mn}/dt \sim 0$ for $\omega_{mn} \ll \Omega_{\rm s}$. The missing $Q^{\prime}_{\rm s\ eq}$ values in Table~\ref{table_tidal_freq_param} correspond to synchronous systems,  systems with $\Omega_{0} < \Omega_{\rm s}$, or without significant tidal interactions, that is, $a/R_{\rm s} > 10$ in our sample. The missing values of $\tau_{\rm spin \max}$ or $\tau_{\rm spin \ ne}$ correspond to systems where such quantities are not defined because the system is synchronous, or $p\Theta  < \kappa_{\rm th} \omega_{mn}^{2}/\Omega_{\rm s}$, or $p \Theta < 0$. The tidal timescale due to the autoresonant oscillations, $\tau_{\rm tide}$, is always much longer than the main-sequence lifetime of the systems because of the oscillations  of the phase lag angle $\psi$ that reduce the average tidal dissipation inside the star (see Figs.~\ref{fig_corot-4}-\ref{fig_CoRoT-6}, lower panels). On the other hand, previous estimates of the same tidal timescale for CoRoT-4, $\tau$~Boo, and WASP-107 in Table~4 of \citet{Lanza22} were shorter than the main-sequence lifetimes of these stars because in that study a constant maximal value $\sin \psi =1$ was assumed. {The confinement of the tidal oscillations to the resonant slender flux tube makes the tidal overlap coefficient as defined in Eq.~\eqref{tidal_overlap_coeff} necessarily small, that is, we confirm the general weakness of the tidal star-planet angular momentum exchanges in our systems  also from the point of view of the more general model cited in Sect.~\ref{general_model}. }

Now we discuss the different kinds of systems by classifying them according to our theory. 

\subsection{Synchronous systems with significant tidal interactions: $\tau$~Bootis and CoRoT-4}

These two systems are characterized by a stellar rotation synchronized with the orbital period, that is, they have virtually reached the final stable state of their tidal evolution. The tidal potential component responsible for the excitation of the autoresonant oscillations is one of the largest among the $\Psi_{lmn}$'s because $l=3$, while $n=m=1$. This makes the threshold values of the frequency drift $\kappa_{\rm th}$ in these two systems five or six orders of magnitude larger than the $\kappa_{\rm exp}$ values expected by considering a tidal modified quality factor $\log_{10} Q^{\prime}_{\rm s}=7.3$ and the wind angular momentum loss rate as given by Eq.~\eqref{angular_mom_loss_rate} (cf. Table~\ref{table_auto_res_param}). Therefore, these two systems can have easily entered the autoresonant regime when they reached the synchronous state during their tidal evolution. 

\begin{figure}
    \centering
    \includegraphics[width=9cm,height=10cm,angle=90,trim=1.9cm 2.0cm 2.0cm 2.0cm,clip]{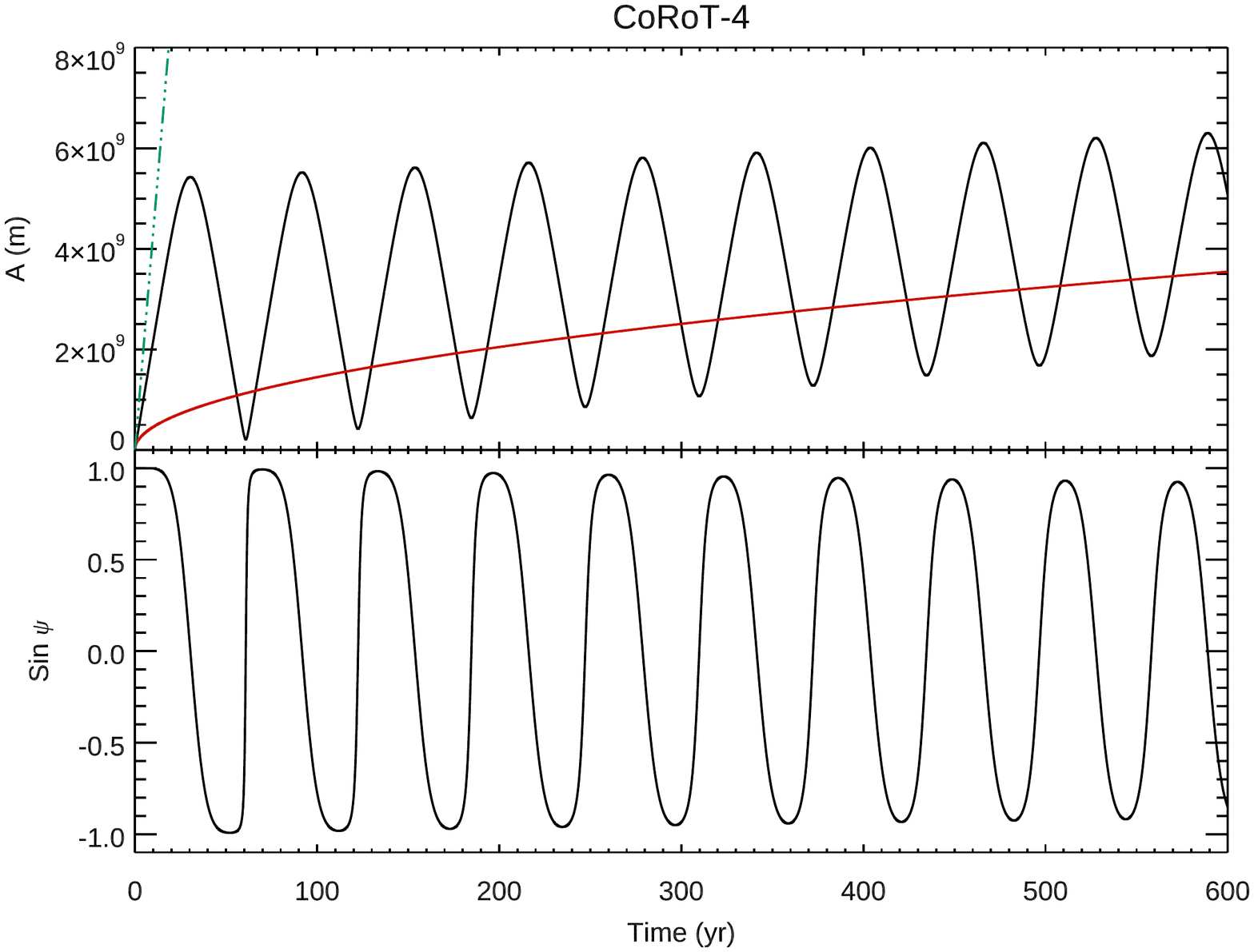}
    \caption{{\em Top panel}: Amplitude of the azimuthal oscillations of the resonant  flux tube vs. the time (black solid line) in CoRoT-4.  The amplitude of the asymptotic solution~\eqref{asymptotic_sol} is plotted as a red solid line for $A_{0} =0$, while the amplitude in the case of a perfect and continuous phase locking ($\sin \psi =1$, constant) is plotted as a green three-dot-dashed line. {\em Lower panel}: the sine of the phase lag angle $\psi$ vs. the time in CoRoT-4. }
    \label{fig_corot-4}
\end{figure}

The evolution of the amplitude $A$ of the azimuthal oscillation $\hat{\xi}_{\phi}$ and  the phase lag $\psi$ for CoRoT-4 can be computed by Eqs.~\eqref{aeq_evol} and~\eqref{psieq_evol} and are plotted in Fig.~\ref{fig_corot-4}. The adopted value of the damping parameter is $2\zeta = 1.5 \times 10^{-15}$~s$^{-1}$, while the value of the frequency drift $\kappa = 10^{3} \kappa_{\rm exp}$ for illustration purposes. The differential equations~\eqref{aeq_evol} and~\eqref{psieq_evol} are numerically integrated by using {\tt ODEINT} \citep{Pressetal02}. The amplitude increases by making oscillations, while its mean becomes more and more comparable with the asymptotic value~\eqref{asymptotic_sol} as expected in an autoresonant regime. Note the very short timescale for the increase of the oscillation amplitude thanks to the  strong  tidal forcing  in such a synchronous system and the value of $\kappa$ remarkably smaller than the threshold $\kappa_{\rm th}$. The maximum amplitude reaches values about one order of magnitude above the estimated $\hat{\xi}_{\phi \max}$ listed in Table~\ref{table_tidal_freq_param} because the equations are integrated without applying any upper bound. The behaviour in the case of $\tau$~Boo is similar, therefore it is not plotted for simplicity. 

The future evolution of the CoRoT-4 and $\tau$~Boo systems is ruled by the loss of angular momentum in their stellar winds. Considering a reduced angular momentum loss rate in the autoresonant regime, both systems can remain into the present synchronous state because $\tau_{\rm red \ AMLR} \sim 2 \times 10^{10}$~yr and $\sim 4.6 \times 10^{9}$~yr for CoRoT-4 and $\tau$~Boo, respectively, when $dL_{\rm w}/dt$ is reduced by a factor of 10 with respect to the values given by Eq.~\eqref{angular_mom_loss_rate}. Both timescales are longer than or comparable with the main-sequence lifetimes of those systems having F-type hosts ($M_{\rm s} \ga (1.2-1.4)$~M$_{\odot}$), therefore, they are expected to maintain their synchronous state for very long time. A reduced angular momentum loss  is crucial in maintaining the synchronization of $\tau$~Boo because it would be difficult to explain with a  normal $dL_{\rm w}/dt$ value given that the timescale for the evolution of the stellar spin would be of $\sim 5 \times 10^{8}$~yr in that case. 


\subsection{Non-synchronous systems with significant tidal interactions: Kepler-13A, Kepler-17 and CoRoT-2}
\label{non-synchronous_systems}

In these systems, the value of $a/R_{\rm s} \la 6.5$ and $m_{\rm p} \ga 2.5$~M$_{\rm J}$ indicating  a remarkable tidal interaction. CoRoT-2 and Kepler-17 are examples of systems where $\Omega_{0} > \Omega_{\rm s}$, $\Theta> 0$,  and $p \equiv m-n > 0$, so that the tidal torque acting on the star can counterbalance the wind braking, while Kepler-13A is an example of a system with $\Omega_{0} < \Omega_{\rm s}$, $\Theta < 0$, and $ p < 0$ where both tides and stellar wind contribute to brake stellar rotation. 

\begin{figure}
    \centering
    \includegraphics[width=8cm,height=10cm,angle=90,trim=1.9cm 2.0cm 2.0cm 2.0cm,clip]{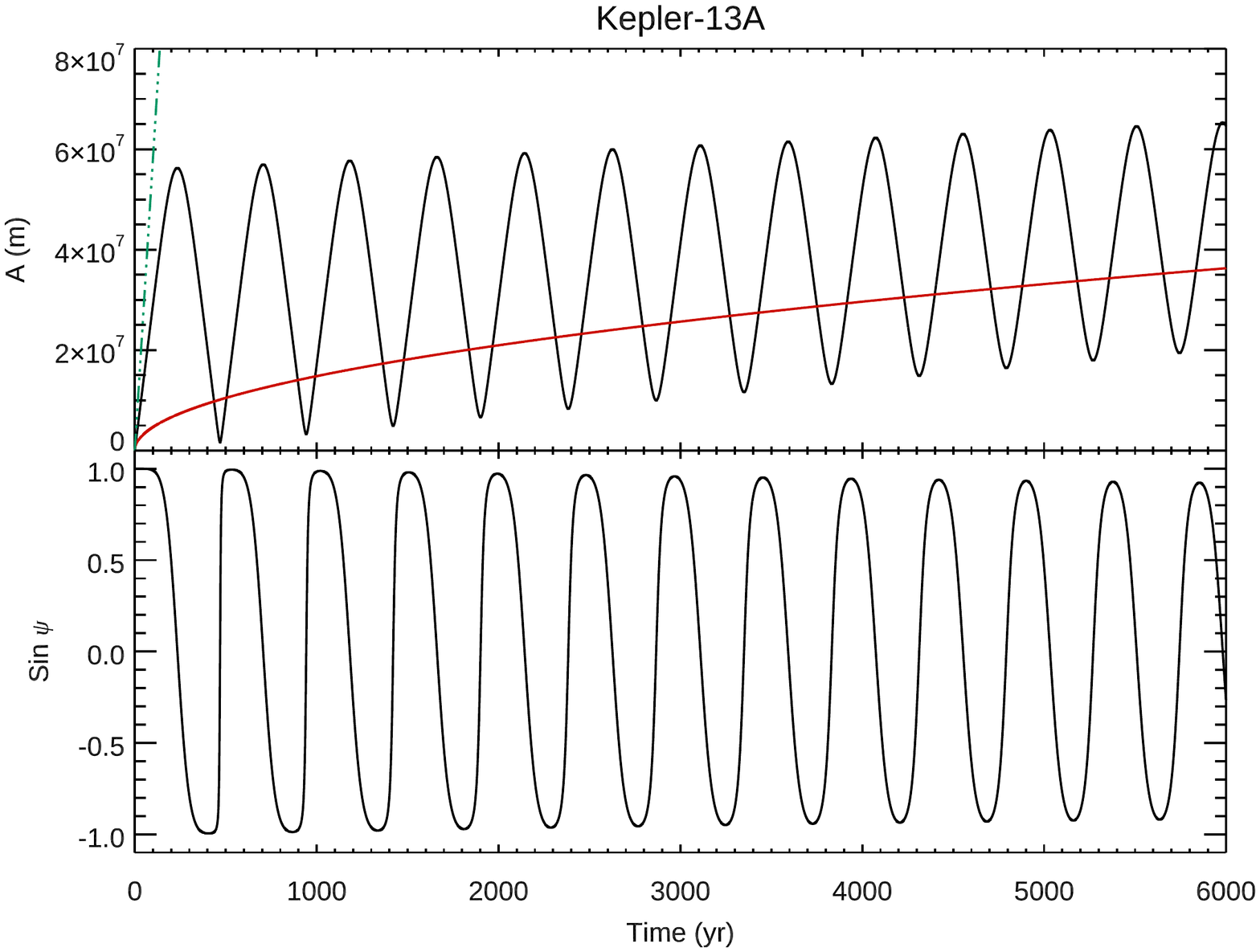}
    \caption{Same as Fig.~\ref{fig_corot-4}, but for Kepler-13A. }
    \label{fig_Kepler-13A}
\end{figure}

In Kepler-13A, the predicted value of the frequency drift $\kappa_{\rm exp}$ computed on the basis of our model is $\sim 4$ times smaller than the threshold value $\kappa_{\rm th}$ for entering autoresonance (cf. Table~\ref{table_auto_res_param}). Given the present uncertainties on the tidal and wind braking processes, we regard this system as capable of entering autoresonance. It can last for all the main-sequence lifetime of the system given that $\tau_{\rm red \, AMLR} \sim 10^{9}$~yr. In this particular system, the host star has a mass of $\sim 1.7$~M$_{\odot}$ making the angular momentum loss intrinsically weak as observed in similar early F-type  stars. Assuming a tidal quality factor $Q^{\prime}_{\rm s} \sim 2 \times 10^{7}$, in agreement with the estimate by  \citet{Lanzaetal11b} for the F-type star CoRoT-11, we obtain a tidal braking timescale of the stellar spin of $|\Theta|^{-1} \sim 0.7$~Gyr in Table~\ref{table_tidal_freq_param}. Nevertheless, CoRoT-11 has a mass of $1.2$~M$_{\odot}$, thus $Q^{\prime}_{\rm s}$ is probably larger for the more massive host Kepler-13A making such a timescale correspondingly longer. In conclusion, a slow evolution away from the present commensurability state is likely to occur in this system, independently of any decrease in the angular momentum loss rate due to the autoresonant phenomenon. 
As an illustration of the autoresonant regime in Kepler-13A, we plot in Fig.~\ref{fig_Kepler-13A} the amplitude and the phase lag computed for $\kappa=0.03 \kappa_{\rm exp}$ and $2\zeta = 1.5 \times 10^{-15}$~s$^{-1}$ to warrant a faster excitation of the oscillations. 

In CoRoT-2, the estimated value of the tidal frequency drift $\kappa_{\rm exp}$ is about $40$ times larger than the threshold value $\kappa_{\rm th}$, while the discrepancy is only within a factor of two in the case of Kepler-17 (cf. Table~\ref{table_auto_res_param}). Therefore, the triggering of the autoresonance requires a temporarily remarkable reduction of the spin evolution in CoRoT-2  and/or a balance between the tidal torque and the wind angular momentum loss in this host in order to reduce the drift of the frequency $\omega_{mn}$ below the threshold. A balance between the tidal torque and the intense wind produced by the high level of magnetic activity of CoRoT-2 requires $Q^{\prime}_{\rm s} \sim 10^{7}$,  assuming a reduction of the braking efficiency  by a factor of ten due to its close-by massive planet (cf. Table~\ref{table_tidal_freq_param}). Such a value of $Q^{\prime}_{\rm s}$ is rather difficult to justify with the equilibrium tide only, but it cannot be excluded given the present uncertainties in the tidal theory in fast-rotating stars \citep[e.g.][]{Barker20,Wei22}. On the other hand, Kepler-17 could have entered the autoresonant regime with a value of $Q^{\prime}_{\rm s}$ as large as $6 \times 10^{8}$ when we assume a reduction by a factor of ten in the angular momentum loss rate  with respect to Eq.~\eqref{angular_mom_loss_rate} (cf. Table~\ref{table_tidal_freq_param}). Such a value of $Q^{\prime}_{\rm s}$ is perfectly in line with the expectations based on the theory by \citet{OgilvieLin07} and the results of \citet{CollierCameronJardine18} for equilibrium tide.

\begin{figure}
    \centering
    \includegraphics[width=8cm,height=10cm,angle=90,trim=1.9cm 2.0cm 2.0cm 2.0cm,clip]{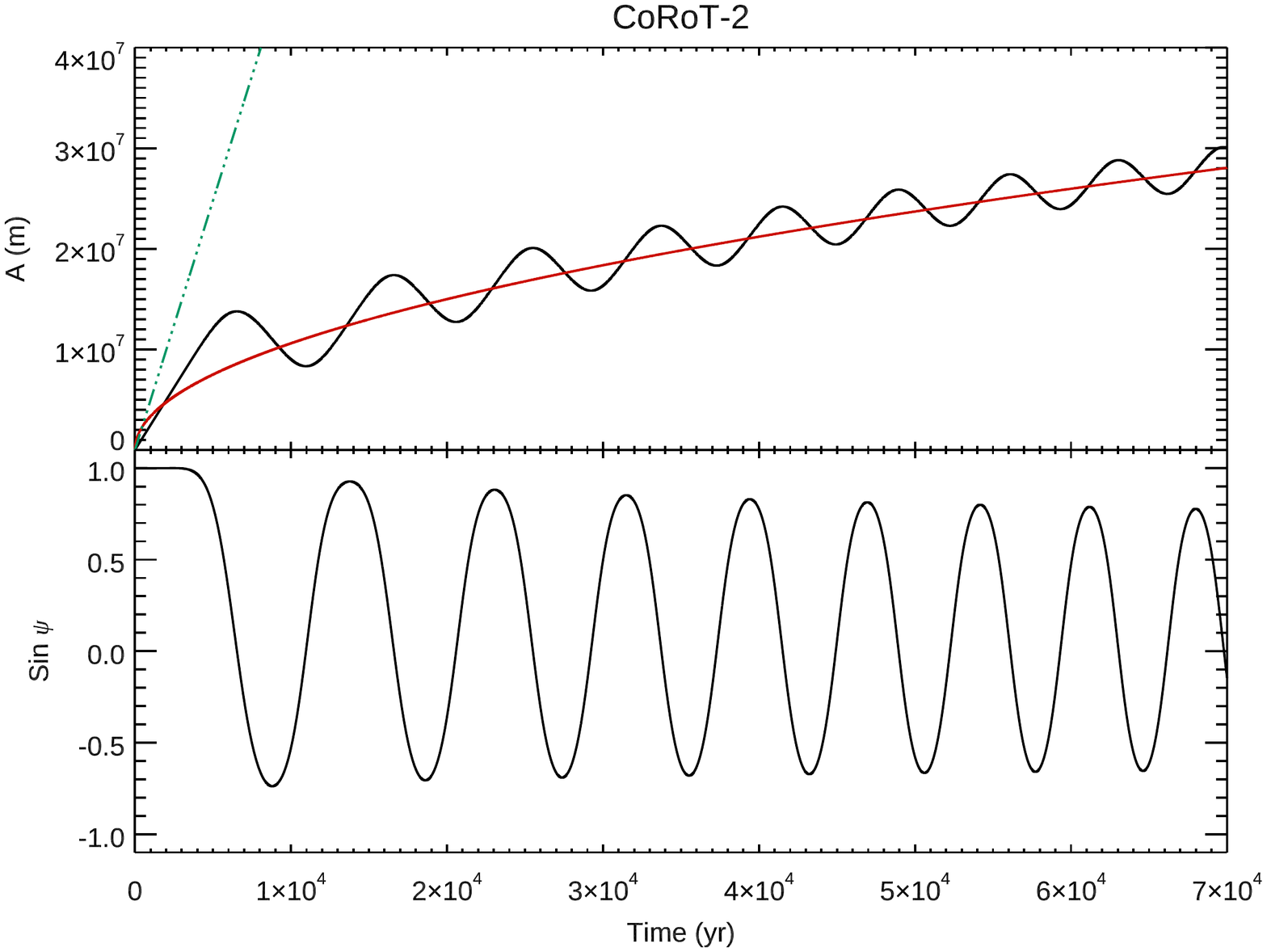}
    \caption{Same as Fig.~\ref{fig_corot-4}, but for  CoRoT-2. }
    \label{fig_corot-2}
\end{figure}

\begin{figure}
    \centering
    \includegraphics[width=8cm,height=10cm,angle=90,trim=1.9cm 2.0cm 2.0cm 2.0cm,clip]{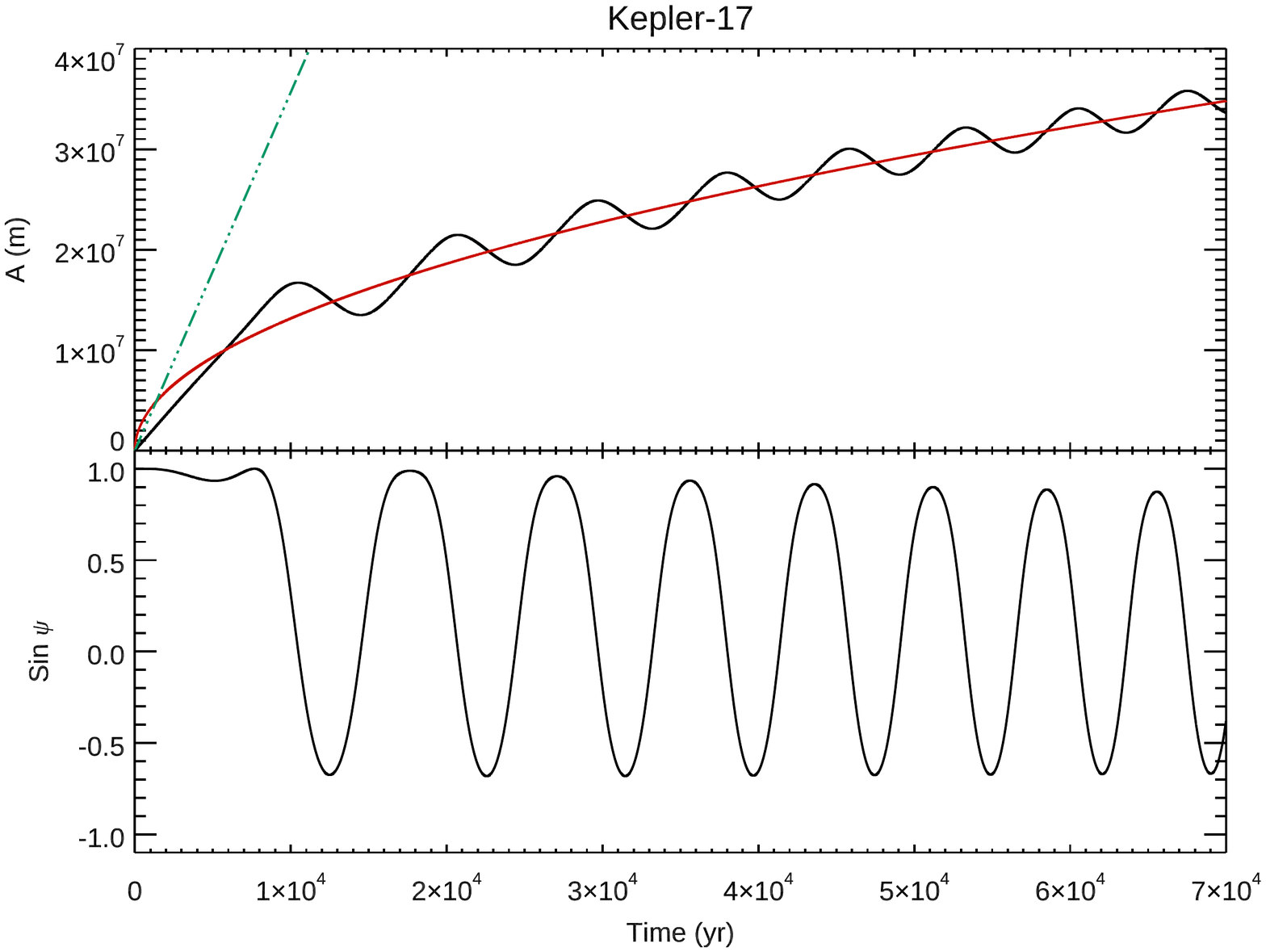}
    \caption{Same as Fig.~\ref{fig_corot-4}, but for Kepler-17. }
    \label{fig_Kepler-17}
\end{figure}

The evolution of the amplitude $A$ and  the phase lag $\psi$ of the autoresonant oscillations are illustrated in Figs.~\ref{fig_corot-2} and~\ref{fig_Kepler-17} for CoRoT-2 and Kepler-17, respectively. For illustrative purposes, we adopted $\kappa$ values corresponding to a reduction by a factor of 200 and 0.3 with respect to the values of $\kappa_{\rm exp}$ listed in Table~\ref{table_auto_res_param}, respectively, $2\zeta = 1.5 \times 10^{-15}$~s$^{-1}$, and do not include an upper bound on the amplitude.

Our model predicts the possibility of a long duration of the autoresonant regime and of the $P_{\rm rot}$-$P_{\rm orb}$ commensurability in both systems, if the tidal torque is capable of balancing the wind torque, due to the reduced angular momentum loss rate in the wind. While a detailed model of the rotational evolution of CoRoT-2 is beyond the scope of the present work, an application of our simplified model in Sect.~\ref{evolution_of_omega_mn} shows that a value of $  3.6 \times 10^{9} \la \tau_{\rm spin} \la 5.3 \times 10^{9}$~yr may keep the system in autoresonance all along its main-sequence lifetime. The corresponding reduction in the angular momentum loss rate with respect to the value computed with Eq.~\ref{angular_mom_loss_rate} is by a factor of $\sim 14-20$. This result gives support to  \citet{PoppenhaegerWolk14} who suggest that CoRoT-2 has an age older than $4-5$~Gyr, at variance with its fast rotation indicating an age $<0.5$~Gyr by comparison with similar single stars  in open clusters. 

In the case of Kepler-17,  $ \tau_{\rm spin} \ga 4.5 \times 10^{9}$~yr is predicted by our model. It is longer by a factor of only $\sim 1.5$  than predicted by  Eq.~\eqref{angular_mom_loss_rate}, implying a modest reduction in the angular momentum loss rate, maybe because Kepler-17 is rotating about three times slower than CoRoT-2. If $\tau_{\rm spin}$ is close to $\tau_{\rm spin \, ne}\sim 1.7 \times 10^{10}$~yr, the initial commensurability is maintained all along the main-sequence evolution given the long timescale $\Theta^{-1} \sim 15$~Gyr. 

 An important consequence of these results is that gyrochronology may not be applicable to such systems owing to their slowed spin-orbit evolution in the autoresonance regime.

\subsection{Systems with negligible tidal interactions }
\begin{figure}
    \centering
    \includegraphics[width=8cm,height=10cm,angle=90,trim=1.9cm 2.0cm 2.0cm 2.0cm,clip]{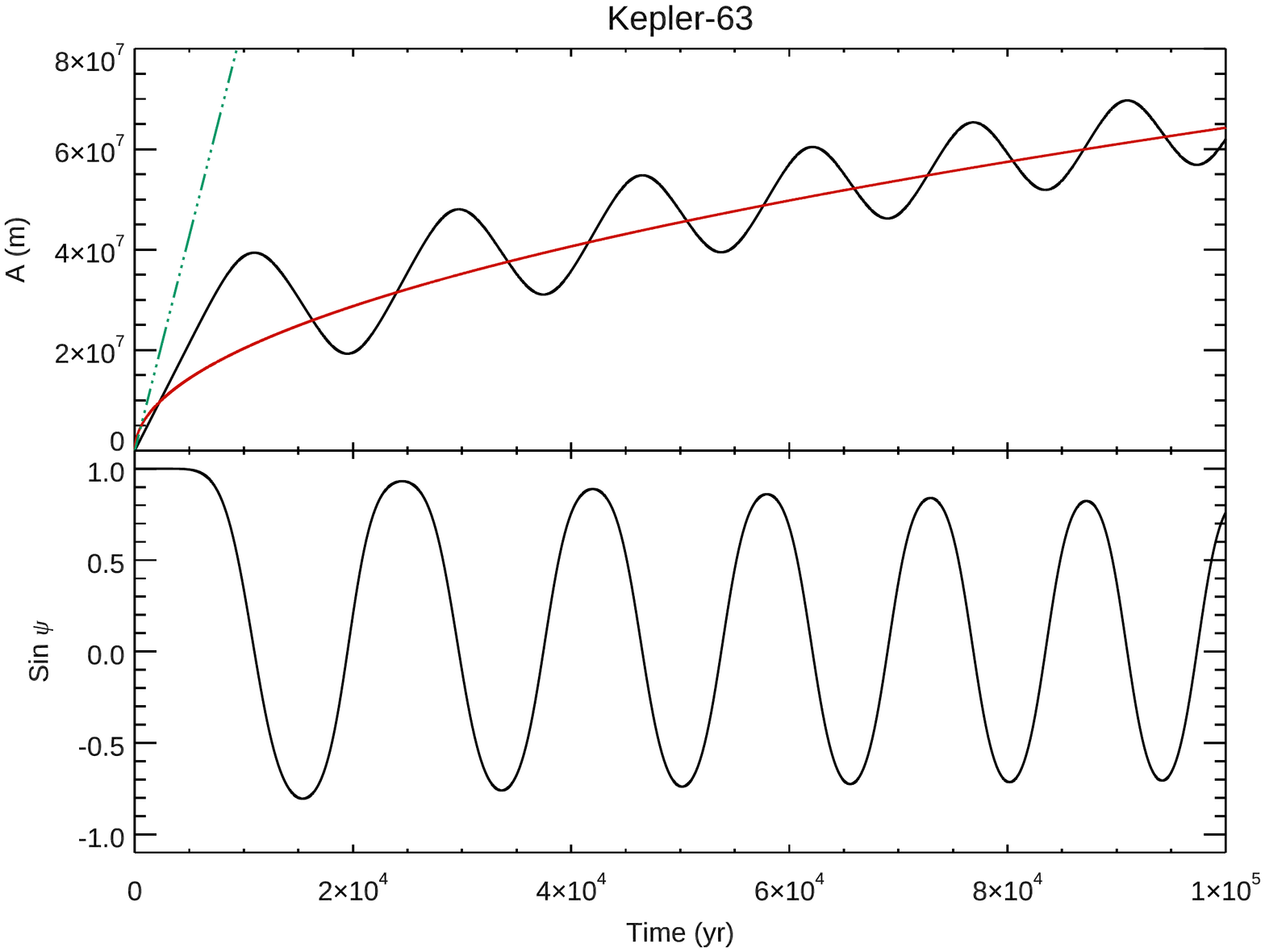}
    \caption{Same as Fig.~\ref{fig_corot-4}, but for Kepler-63. }
    \label{fig_Kepler-63}
\end{figure}

Systems with a negligible tidal interaction are characterized by $a/R_{\rm s} > 10$ and include CoRoT-6, Kepler-63, HAT-P-11, WASP-107, and AU~Mic. Kepler-63 and WASP-107 have an expected frequency drift $\kappa_{\rm exp}$ comparable with and much smaller than their threshold values $\kappa_{\rm th}$, respectively, thus those systems can be expected to have entered the autoresonance phase. For illustration purpose, we plot the amplitude and phase lag in autoresonance for Kepler-63 in Fig.~\ref{fig_Kepler-63} for $\kappa= 0.005 \kappa_{\rm exp}$ and $2\zeta = 1.5 \times 10^{-15}$~s$^{-1}$. 

\begin{figure}
    \centering
    \includegraphics[width=8cm,height=10cm,angle=90,trim=1.9cm 2.0cm 2.0cm 2.0cm,clip]{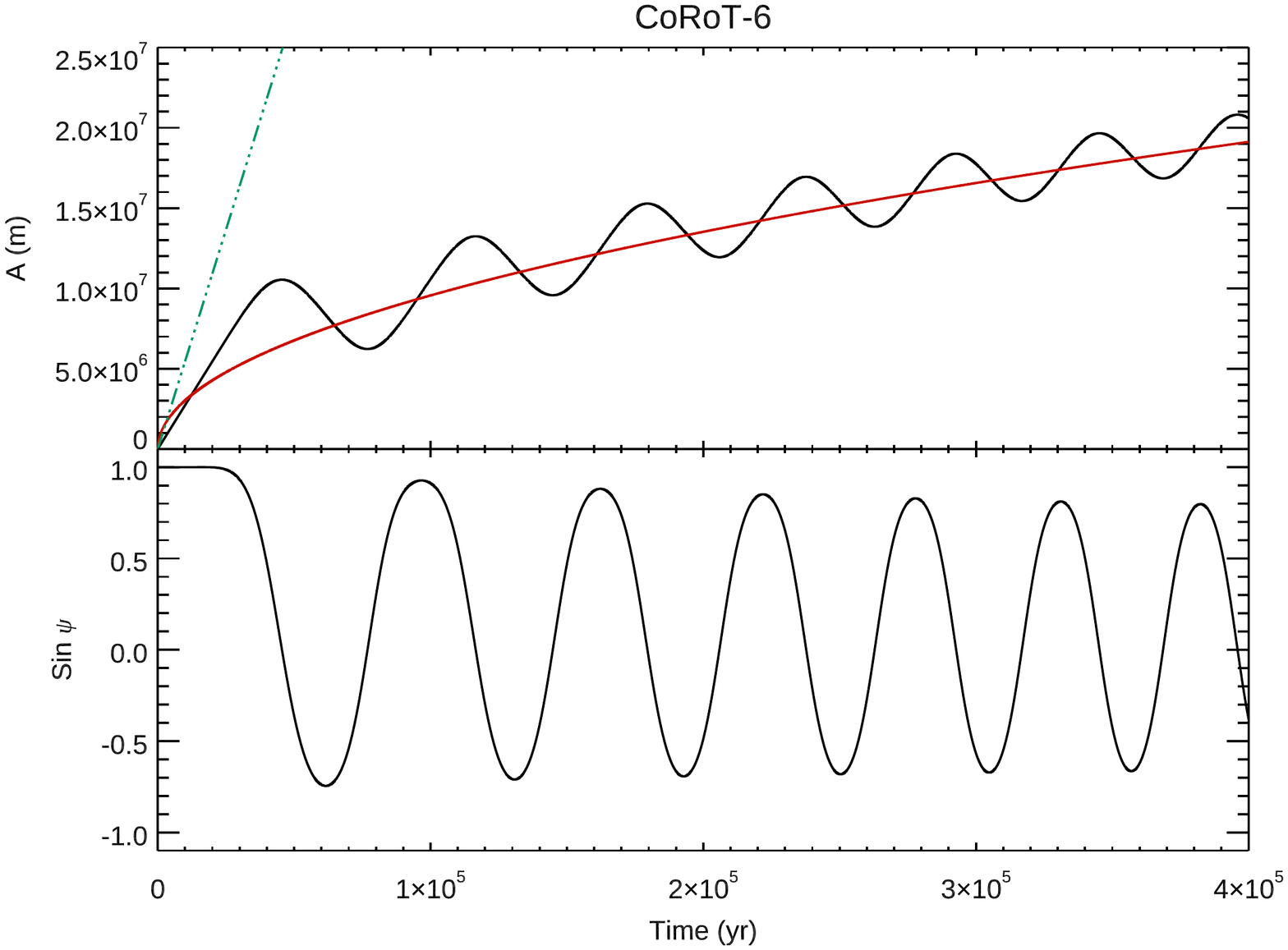}
    \caption{Same as Fig.~\ref{fig_corot-4}, but for CoRoT-6. }
    \label{fig_CoRoT-6}
\end{figure}

For the other systems, the estimated values of $\kappa$ are larger than the thresold values by factors of $\sim 400$ for CoRoT-6, $\sim 5000$ for HAT-P-11, and $8 \times 10^{8}$ for AU~Mic, the latter two representing cases that cannot be accounted for by the model proposed by \citet{Lanza22}. A phase of stalled rotational evolution could have allowed CoRoT-6 to enter autoresonance because a constant stellar rotation lasting for  $\la 10^{5}$~yr is enough to trigger autoresonance in such a system as illustrated in Fig.~\ref{fig_CoRoT-6} where we plot amplitude and phase lag for $\kappa = 5 \times 10^{-4} \kappa_{\rm exp}$ and $2\zeta = 1.5 \times 10^{-15}$~s$^{-1}$.

In the case of HAT-P-11 and AU~Mic, the maximum amplitude of the azimuthal resonant oscillations is only $\hat{\xi}_{\phi \max} \sim 150-300$~m, that is, at least four orders of magnitude smaller than for the other systems listed in Table~\ref{table_tidal_freq_param}. This is a consequence of the weakness of the  tidal potential in these two systems owing to $a/R_{\rm s} > 15$ and $m_{\rm p} < 0.1$~M$_{\rm J}$. Such small oscillation amplitudes are unlikely to significantly perturb the neighbour magnetic flux tubes responsible for stellar photospheric activity, thus indicating that also the present model may not explain the observed commensurability in HAT-P-11 and AU~Mic. Therefore, other models should be considered to account for a potential influence of the close-by planets on the activity of these hosts  \cite[e.g.,][]{Lanza08,Abreuetal12,Stefanietal18}\footnote{For a different point of view, see, for example, \citet{CameronSchussler13}. }. 

{The possibility that a low-frequency resonant oscillation mode of a star evolves with the orbital and stellar spin frequencies in order to maintain the commensurability condition has been proposed in stellar binaries and has been called resonance locking \citep[see, e.g.,][]{Fuller17,ZanazziWu21}. Its applicability to the star-planet case  remains to be investigated. In the specific case of AU Mic and HAT-P-11,  this mechanism is highly unlikely because of the extremely long tidal evolution timescales in those systems, much longer than the evolution timescales of  their stellar structures, while the latter are required to be larger than the former in order to make the mechanism to work \citep[cf. Eq. 34 in][]{ZanazziWu21}.} 

The duration of the autoresonant regime in the systems considered in this subsection is predicted to be comparable with their main-sequence lifetime because of the weakness of the tidal interactions making the timescales $|\Theta|^{-1}$ much longer than their host star nuclear timescales. With the possible exception of WASP-107, the minimum spin evolution timescales $\tau_{\rm spin \min}$ are also very long because of the reduction in the angular momentum loss rate in autoresonance. Therefore, no significant evolution is predicted in their $P_{\rm rot}/P_{\rm orb}$ ratio during their main-sequence lifetime according to our model, except for the pre-main-sequence star AU~Mic that is still contracting to reach the zero-age main sequence. 

\section{Discussion and conclusions}
\label{discussion}
We introduced a model that can interpret the close spin-orbit commensurability observed in several planetary systems in the framework of the approach proposed by \citet{Lanza22} that considers the tidal excitation of resonant oscillations in a magnetic slender flux tube stored into the overshoot layer of their host stars. Specifically, our model accounts for the excitation and the amplification of such oscillations in spite of the weakness of the tidal potential by considering that the system can enter an autoresonant regime. In such a regime, the flux tube oscillations can be kept in phase with the oscillations of a weak tidal potential component for a sufficiently long time interval as needed to amplify them. A consequence of the  oscillations of such a resonant flux tubes is the triggering of the undulatory instability in other flux tubes stored in the overshoot layer having a stronger magnetic field that makes them emerge to the photosphere where they form a pattern characterized by a high-$l$ coronal field topology. This leads to a remarkable reduction in the angular momentum loss rate that may keep these systems in the autoresonant regime for timescales comparable with their main-sequence lifetime. 

We analysed the same ten systems considered by \citet{Lanza22} finding that autoresonance is a viable amplification mechanism for eight of them with the exception of HAT-P-11 and AU~Mic for which the mechanism produces a maximum oscillation amplitude at least four orders of magnitudes smaller than in the other systems. 

{It is important to stress that the introduced model is  very idealized  being based on an individual resonant magnetic flux tube in isolation (cf. Sect.~\ref{general_model}). The configuration of the magnetic field in the overshoot layers of late-type stars is largely unknown, thus it is not possible to assess the relevance of this simplifying assumption in real stars. A future direction of study can be the consideration of low-frequency oscillations in a magnetized layer as proposed by, e.g., \citet{Zaqarashvili18}, to see whether they can be tidally excited in the case of a close spin-orbit commensurability. 

Most of the previous studies on the effect of magnetic fields on dynamical tides focused on the dissipation of the tidal energy  that was  found to be generally increased  because of the added effect of Ohmic dissipation \citep[cf.][]{Wei16,Wei18,LinOgilvie18}. On the other hand, \citet{Astouletal19} considered the role of magnetic fields in the excitation of dynamical tides and found that it can become relevant in the case of a commensurability between the stellar spin and the orbital mean motion of a  planet on an oblique or eccentric orbit. Such a result, although obtained from an order-of-magnitude comparison between the Lorentz force and the hydrodynamical effects in the excitation of the dynamical tides, provides support for the proposal by \citet{Lanza22} and the present idealized excitation model. In other words, the possible tidal excitation of magnetostrophic oscillations in stars hosting planets in spin-orbit commensurability is worth of further investigation.}

The present study also suggests that gyrochronology cannot be trusted in the case of stars with close-by massive planets, notably when they show a commensurability between their rotation and orbital periods. 

Our treatment allows us to compute the tidal dissipation inside stars in the autoresonant regime finding values remarkably smaller than those predicted by \citet{Lanza22} with the simplifying assumption of a perfect and continuous phase locking between the oscillations and the forcing tidal potential. By computing the actual phase lag between forcing and autoresonant oscillations, we find that the average torque associated with the dissipation of the flux tube oscillations is in general negligible in comparison with those of the other dynamical tides proposed to operate in late-type planetary hosts \citep{OgilvieLin07,Mathis15,Barker20}. 

The fraction of close-by planetary systems displaying a spin-orbit commensurability can be roughly estimated by considering, for example, the compilation by \citet{Albrechtetal21} where the rotation periods for a total of 51 host stars are reported. 
Those systems were selected  because they have a transiting close-by planet with a measurement of its projected orbit obliquity, a property that should not, to our knowledge, be correlated with the presence of a spin-orbit commensurability. Among those stars, we find seven of our systems, that is, $\sim 15$\% of their sample. Assuming that all those cases are not due to a chance coincidence, this would correspond to a duration of the commensurability regime of at least  $\sim 1.5$~Gyr, assuming an average main-sequence lifetime of $\sim 10$~Gyr and that all the systems sooner or later enter such a regime. On the other hand, if only some fraction of the systems experience a phase of commensurability during their evolution, the duration of this phase can be longer as suggested in Sect.~\ref{evolution_of_omega_mn}. 


\begin{acknowledgements}
      The author is grateful to an anonymous Referee for pointing out the limitations of the present idealized model and the importance of discussing them when applications to real stars are considered. 
      This work is dedicated to the Ukrainian people, victims of a barbaric war of aggression that is bringing death, deep suffering, and destruction in their wonderful country. This investigation has been supported by the Italian National Institute for Astrophysics through the INAF-PRIN project PLATEA (P.I. Dr. Silvano Desidera). 
\end{acknowledgements}

\appendix
\section{Non-linear dependence of the oscillation frequency due to differential rotation}
\label{non-linear_dep_and_dr}
The magnetic field perturbation due to the radial oscillations of a slender flux tube can be computed by considering the ideal induction equation, that is, neglecting magnetic diffusion because its timescale is on the order of $10^{7}$~yr in our model, much longer than the period of the oscillations. The unperturbed field ${\vec B} = B \, \hat{\vec e}_{\phi}$ of the flux tube is purely azimuthal, but the oscillations  induce a radial perturbation given by \citep[see Sect.~3.9.1 of][]{Lanza22}
\begin{equation}
    \delta B_{r} = j m \frac{B \xi_{r}}{r_{0}},
\end{equation}
where $j=\sqrt{-1}$, $m$ is the azimuthal order of the oscillation mode of radial displacement $\xi_{r}$, and $r_{0}$ the radius of the unperturbed flux tube. If there is a radial differential rotation in the layer where the flux tube is stored, it  induces a perturbation of the azimuthal field $B$ according to the induction equation
\begin{equation}
    \frac{\partial B}{\partial t} = r_{0}  \Omega_{\rm s}^{\prime} (\delta B_{r}), 
    \label{azim_induct}
\end{equation}
where $\Omega_{\rm s}^{\prime} = \partial \Omega_{\rm s}/\partial r$ is the radial gradient of the stellar angular velocity $\Omega_{\rm s}$ and the flux tube is assumed to be in the equatorial plane. Equation~\eqref{azim_induct} can be integrated with respect to the time by dividing it by $j \omega_{0}$, where $\omega_{0}$ is the unperturbed frequency of the oscillation mode, yielding
\begin{equation}
    \delta B = \left(\frac{m}{\omega_{0}}\right)\,  B\, \Omega_{\rm s}^{\prime}\, \xi_{r}. 
\end{equation}
The time average of the variation in the square of the magnetic field is $(\delta B)^{2}$ because the first-order term $B \delta B$ has a zero average over one oscillation period. The corresponding variation in the frequency of the oscillation mode is (cf. Sect.~\ref{oscill_freq_sect}) 
\begin{equation}
    \omega = \omega_{0} \left( 1 + \gamma_{\rm r} \, \hat{\xi}_{r}^2 \right), 
\end{equation}
where 
\begin{equation}
    \gamma_{\rm r} \equiv \frac{1}{4} \left( \frac{m}{\omega_{0}} \Omega_{\rm s}^{\prime}\right)^{2}. 
\end{equation}
Considering $\omega_{0} \sim 10^{-4}\, \Omega_{\rm s}$, $m=8$, and a rather extreme $\Omega_{\rm s}^{\prime} \sim 10 \, \Omega_{\rm s}/R_{\rm s}$, where $R_{\rm s}$ is the radius of the star assumed equal to the solar radius, we estimate $\gamma_{\rm r} \sim 3 \times 10^{-7}$~m$^{-2}$ for a Sun-like star. The ratio $\gamma_{\rm r} / \gamma_{0} \sim (\Omega^{\prime} r_{0} / \omega_{0})^{2} \ll 1$ (cf. Sect.~\ref{oscill_freq_sect}). Therefore, the present non-linearity in the oscillation frequency of the flux tube is negligible in comparison with that introduced in Sect.~\ref{oscill_freq_sect}, {\em a fortiori} given that $\xi_{r} \ll \xi_{\phi}$. 

\section{Multiple timescale method: application to the forced Duffing oscillator}
\label{autoresonance_eqs}

We follow the approach introduced in \citet{Arovas20}; other more general approaches may be found in, for example, \citet{Chirikov79} or \citet{Friedland01}. We consider an oscillator described by the equation of motion
\begin{equation}
    \ddot{x} + x= \epsilon h(x, \dot{x}),
    \label{a2_eq1}
\end{equation}
where $\epsilon \ll 1$ is a small parameter, $h$ is a non-linear function of the oscillator position $x$ and its velocity $\dot{x}$, while the time unit is the inverse of the unperturbed oscillator frequency as in Sect.~\ref{autoresonance_section}; the time derivative is indicated by a dot over the symbol of a variable. We limit ourselves to the first-order theory and define a normal time $T_{0}$ and a slow time $T_{1} \equiv \epsilon T_{0}$, such that the time derivative is given by
\begin{equation}
    \frac{\partial}{\partial t^{\prime}} \equiv \frac{\partial}{\partial T_{0}} + \epsilon \frac{\partial}{\partial T_{1}}
\end{equation}
and we expand 
\begin{equation}
    x = x_{0} + \epsilon x_{1}. 
\end{equation}
Equation~\eqref{a2_eq1} becomes
\begin{eqnarray}
    \left[ 1 + \left( \frac{\partial}{\partial T_{0}} + \epsilon \frac{\partial}{\partial T_{1}} \right)^{2}\right] (x_{0} + \epsilon x_{1})  =   \nonumber \\
= \epsilon h \left[ (x_{0} +\epsilon x_{1}), 
    \left(\frac{\partial}{\partial T_{0}} + \epsilon \frac{\partial}{\partial T_{1}} \right) (x_{0} + \epsilon x_{1}) \right]. 
    \label{a2_eq2} 
\end{eqnarray}
Considering only the terms up to the first order in $\epsilon$, we find
\begin{eqnarray}
   \left( \frac{\partial}{\partial T_{0}} + \epsilon \frac{\partial}{\partial T_{1}} \right) (x_{0} + \epsilon x_{1}) & = & \nonumber \\
   \frac{\partial x_{0}}{\partial T_{0}} + \epsilon \left( \frac{\partial x_{1}}{\partial T_{0}} + \frac{\partial x_{0}}{\partial T_{1}} \right) \mbox { so that}\\
   \left( \frac{\partial}{\partial T_{0}} + \epsilon \frac{\partial}{\partial T_{1}} \right)^{2} (x_{0} + \epsilon x_{1}) = \nonumber \\
   \frac{\partial^{2} x_{0}}{\partial T_{0}^{2}} + 2\epsilon \frac{\partial^{2} x_{0}}{\partial T_{0} \partial T_{1}} +\epsilon \frac{\partial^{2} x_{1}}{\partial T_{0}^{2}}. 
\end{eqnarray}
Substituting such expressions into Eq.~\eqref{a2_eq2} and separating the terms of zero order in $\epsilon$ from those of the first order, we obtain
\begin{eqnarray}
   \frac{\partial^{2} x_{0}}{\partial T_{0}^{2}} + x_{0} = 0,  \label{a2_eq3}\\
   \frac{\partial^{2} x_{1}}{\partial T_{0}^{2}} + x_{1} = -2 \frac{\partial^{2} x_{0}}{\partial T_{0} \partial T_{1}} + h \left( x_{0}, \frac{\partial x_{0}}{\partial T_{0}} \right).  
   \label{a2_eq4}
\end{eqnarray}
The solution of Eq.~\eqref{a2_eq3} is 
\begin{equation}
    x_{0} = A \cos (T_{0} + \phi_{\rm A}),
\end{equation}
where the amplitude $A$ and the initial phase $\phi_{\rm A}$ are both functions of the slow time $T_{1}$, that is, they vary very little over one period of the oscillation of $x_{0}$. If we define $\theta \equiv T_{0} + \phi_{\rm A}$, then $x_{0} =A \cos \theta$, while $\partial /\partial T_{0} = \partial / \partial \theta$, and Eq.~\eqref{a2_eq4} becomes
\begin{eqnarray}
\frac{\partial^{2} x_{1}}{\partial T_{0}^{2}} + x_{1}  = -2 \frac{\partial }{\partial T_{1}} \left( - A \sin \theta \right) + h \left( A \cos \theta, -A \sin \theta \right) \mbox{ or} \nonumber \\
 \frac{\partial^{2} x_{1}}{\partial T_{0}^{2}} + x_{1}  =  2 \frac{\partial A}{\partial  T_{1}} \sin \theta  \nonumber \\ 
    + 2 A \frac{\partial \phi_{\rm A}}{\partial T_{1}} \cos \theta + h ( A\cos \theta , -A \sin \theta).  \label{a2_eq12}
\end{eqnarray}
The motion of the oscillator is periodic, therefore, $h$ takes the same value when $\theta$ is increased by $2\pi$. This gives the possibility of expanding $h$ into a Fourier series as
\begin{equation}
    h(A \cos \theta, -A \sin \theta ) = \sum_{k=1}^{\infty} \alpha_{k} \sin (k \theta) + \sum_{k=0}^{\infty} \beta_{k} \cos (k \theta).
\end{equation}
Similarly, we expand $x_{1}$ into a Fourier series as 
\begin{equation}
    x_{1} = \sum_{k=1}^{\infty} a_{k} \sin (k \theta) + \sum_{k=0}^{\infty }b_{k} \cos (k \theta). 
\end{equation}
Substituing the expansions into Eq.~\eqref{a2_eq12} and taking into account that $A$ and $\phi_{\rm A}$ depend on $T_{1}$, we have 
\begin{eqnarray}
    \sum_{k} \left[ (1-k^{2})\, a_{k} \sin (k\theta) + (1-k^{2})\, b_{k} \cos (k\theta) \right] = \nonumber \\
    2 \frac{\partial A}{\partial T_{1}} \sin \theta + 2A \frac{\partial \phi_{\rm A}}{\partial T_{1}} \cos \theta + \sum_{k} \left [\alpha_{k} \sin (k \theta) + \beta_{k} \cos (k \theta) \right]. 
    \label{a2_eq5}
\end{eqnarray}
Equation~\eqref{a2_eq5} can be satisfied for different values of the integer $k$  by taking into account the orthogonality of the functions $\sin (k\theta)$ and $\cos (k \theta)$. For $k=1$, we find 
\begin{eqnarray}
    2 \frac{\partial A}{\partial T_{1}} + \alpha_{1} (A) = 0, \label{a2_eq6}\\
    2 A \frac{\partial \phi_{\rm A}}{\partial T_{1}} + \beta_{1} (A) = 0 \label{a2_eq7}, 
\end{eqnarray}
while for $k \not= 1$, we have 
\begin{eqnarray}
  a_{k} = \frac{\alpha_{k} (A)}{1- k^{2}},  \\
  b_{k} = \frac{\beta_{k} (A)}{1- k^{2}}. 
\end{eqnarray}
Now, we specialize the above results to the case of a damped Duffing oscillator  by considering an equation of motion of the kind
\begin{equation}
    \ddot{x} + 2 \lambda \dot{x} + x -\gamma_{\rm D} x^{3} = 0,   
    \label{a2_duffing}
\end{equation}
where $2 \lambda$ is the inverse of the damping timescale and $\gamma_{\rm D}$ the coefficient that rules the non-linearity. Equation~\eqref{a2_duffing} is equivalent to Eq.~\eqref{a2_eq1} provided that
\begin{equation}
    \epsilon h(x, \dot{x}) = \gamma_{\rm D} x^{3} -2 \lambda \dot{x}. 
\end{equation}
Since $x_{0} = A \cos \theta$ and  $\cos^{3} \theta = (3/4) \cos \theta + (1/4) \cos (3 \theta)$, we find 
\begin{equation}
    \epsilon h(x, \dot{x}) = 2 \lambda A \sin \theta + \frac{3}{4} \gamma_{\rm D} A^{3} \cos \theta + \frac{1}{4} \gamma_{\rm D} A^{3} \cos (3 \theta). 
\end{equation}
This is equivalent to 
\begin{eqnarray}
   \alpha_{1} (A) = 2 \lambda_{0} A(T_{1}) \label{a2_eq15} \\
   \beta_{1} (A) = \frac{3}{4} \gamma_{0} A^{3} (T_{1}) \label{a2_eq16},
\end{eqnarray}
where $\lambda_{0} \equiv \lambda/\epsilon$, $\gamma_{0} \equiv \gamma_{\rm D}/\epsilon$, and we have explicitly indicated the dependence of $A$ on the slow time $T_{1}$. Therefore, Eqs.~\eqref{a2_eq6} and~\eqref{a2_eq7} become 
\begin{eqnarray}
   \frac{\partial A}{\partial T_{1}} = - \lambda_{0} A (T_{1}), \label{a2_eq8}\\
   \frac{\partial \phi_{\rm A}}{\partial T_{1}} = -\frac{3}{8} \gamma_{0} A^{2} (T_{1}).  \label{a2_eq9} 
\end{eqnarray}
The solution of Eq.~\eqref{a2_eq8} is 
\begin{equation}
    A = A_{0} \exp (-\lambda_{0} T_{1}),
\end{equation}
where $A_{0}$ is the initial value of $A$ at $T_{1} = 0$. Substituting into Eq.~\eqref{a2_eq9}, and integrating with respect to $T_{1}$, we find
\begin{equation}
    \phi_{\rm A} = \phi_{\rm A 0} - \frac{3 \gamma_{0} A_{0}^{2}}{16 \lambda_{0}} \left[ 1- \exp \left( -2\lambda_{0} T_{1} \right) \right],
\end{equation}
where $\phi_{\rm A 0}$ is the initial value of $\phi_{\rm A}$.

Finally, we consider the case of the damped Duffing oscillator with an external cosinusoidal forcing. Its equation of motion is (cf. Sect.~\ref{autoresonance_section})
\begin{equation}
    \ddot{x} + 2 \lambda \dot{x} + x - \gamma_{\rm D} x^{3} = \eta \cos \varphi,
    \label{a2_eq10}
\end{equation}
where 
\begin{equation}
    \varphi = t^{\prime} - \frac{1}{2} \kappa \, t^{\prime 2}
\end{equation}
is the phase of the forcing, the frequency of which is varying on the slow timescale due to the term containing $\kappa$ (see below). Equation~\eqref{a2_eq10} can be recast in the form of Eq.~\eqref{a2_eq1} by defining
\begin{equation}
    \epsilon h (x, \dot{x}, \varphi) \equiv \gamma_{\rm D} x^{3} - 2\lambda \dot{x} + \eta \cos \varphi. 
\end{equation}
The zero-th order solution can be assumed again $x_{0} = A(T_{1}) \cos \left[ T_{0} + \phi_{\rm A}(T_{1}) \right] = A \cos \theta$. We introduce the  angle $\psi$ defined as
\begin{equation}
    \psi \equiv \theta - \varphi + \pi = T_{0} + \phi_{\rm A} - \varphi + \pi. 
\end{equation}
Therefore, substituing into Eq.~\eqref{a2_eq12} and considering that 
\begin{equation}
    \frac{\partial \phi_{\rm A}}{\partial T_{1}} = \frac{\partial}{\partial T_{1}} (\psi + \varphi),  
\end{equation}
we find 
\begin{eqnarray}
\frac{\partial^{2} x_{1}}{\partial T_{0}^{2}} + x_{1} = 2 \frac{\partial A}{\partial T_{1}} \sin  \theta + 2 A \frac{\partial \psi}{\partial T_{1}} \cos \theta   \nonumber \\
+ 2 A \frac{\partial \varphi}{\partial T_{1}} \cos \theta + h (A\cos \theta, -A \sin \theta) - \eta_{0} \cos (\theta - \psi + \pi),    \label{a2_eq14}
\end{eqnarray}
where $\eta_{0} \equiv \eta / \epsilon $. The time derivative of $\varphi$ is
\begin{equation}
    \frac{\partial \varphi}{\partial t^{\prime}} = \frac{\partial \varphi}{\partial T_{0}} + \epsilon \frac{\partial \varphi}{\partial T_{1}}. 
    \label{a2_eq13}
\end{equation}
The left-hand side of Eq.~\eqref{a2_eq13} is $\partial \varphi/\partial t^{\prime} = 1 -\kappa t^{\prime}$. Comparing with the right-hand side and separating the zero-order term from the first-order term, we find 
\begin{eqnarray}
   \frac{\partial  \varphi}{\partial T_{0}} = 1, \\
   \frac{\partial \varphi}{\partial T_{1}} = - \frac{\kappa}{\epsilon} T_{0} = -\kappa_{0} T_{0}, 
\end{eqnarray}
where we defined $\kappa_{0} \equiv \kappa /\epsilon$. Making use of these relationships, developing in a Fourier series $h(A\cos \theta, -A \sin \theta)$, and grouping the terms in $\sin \theta$ and $\cos \theta$ in Eq.~\eqref{a2_eq14}, we find
\begin{eqnarray}
   \frac{\partial^{2} x_{1}}{\partial T_{0}^{2}} + x_{1}= \left( 2 \frac{\partial A}{\partial T_{1}} +\alpha_{1} - \eta_{0} \sin \psi \right) \sin \theta \nonumber \\ 
  + \left( 2A \frac{\partial \psi}{\partial T_{1}} - 2A \kappa_{0} T_{0} + \beta_{1} - \eta_{0} \cos \psi \right) \cos \theta  \nonumber \\ 
   + \sum_{k \not= 1} \left[ \alpha_{k} \sin ( k\theta) + \beta_{k} \cos (k \theta) \right]. 
\end{eqnarray}
Thanks to the orthogonality of the sine and cosine terms in the above equation, substituing the expressions for $\alpha_{1}$ and $\beta_{1}$ from Eqs.~\eqref{a2_eq15} and~\eqref{a2_eq16}, we have
\begin{eqnarray}
    \frac{\partial A}{\partial T_{1}} +  \lambda_{0} A - \frac{1}{2}\eta_{0} \sin \psi = 0, \\
   \frac{\partial \psi}{\partial T_{1}} - \kappa_{0} T_{0} + \frac{3\gamma_{0}}{8} A ^{2} -\frac{\eta_{0}}{2A} \cos \psi = 0. 
\end{eqnarray}
Considering that $\partial/\partial T_{1} = \epsilon^{-1} \partial / \partial T_{0}$, for $T_{1} = \epsilon T_{0}$, and taking into account the definitions of the coefficients $\kappa_{0}$, $\lambda_{0}$, $\gamma_{0}$, and $\eta_{0}$ given above, we finally found
\begin{eqnarray}
   \frac{\partial A}{\partial T_{0}} = - \lambda A + \frac{1}{2} \eta \sin \psi,  \\
   \frac{\partial \psi}{\partial T_{0}} = \kappa T_{0} - \frac{3 \gamma_{\rm D}}{8} A^{2} + \frac{\eta}{2A} \cos \psi, 
\end{eqnarray}
that are Eqs.~\eqref{aeq_evol} and~\eqref{psieq_evol}, respectively, if we identify the normal time $T_{0}$ with the time $t^{\prime}$ as defined in Sect.~\ref{autoresonance_section}. 
\end{document}